\documentclass[%
 reprint,
 amsmath,amssymb,
 aps,
]{revtex4-1}

\usepackage[utf8]{inputenc}
\usepackage[english]{babel}

\usepackage{graphicx}
\usepackage{dcolumn}
\usepackage{bm}
\usepackage{float}

\setcitestyle{authoryear,open={(},close={)}}

\begin{document}

\preprint{APS/123-QED}

\title{Targeted Exploration of Dark Photon Parameter Space at Mu3e}

\author{I. Fernando}%
\thanks{These authors contributed equally.}
\author{S. Hackett}%
\thanks{These authors contributed equally.}
\author{A. Boulton McKeehan}
\thanks{These authors contributed equally.}
\affiliation{%
Stanford University Physics Department
}%

\date{\today}

\begin{abstract}
In 2016, the Atomki collaboration discovered an anomaly in the decay of beryllium excited states that matched theoretical predictions for an unknown force-carrying boson. We suggest a search for such a hypothesized `dark photon' over the same parameter space to be carried out at the Mu3e detector. We discuss the mechanisms by which a dark photon may be observed, and outline the details of this detector experiment. We extrapolate on this groundwork to analyze projections for the sensitivity of the experiment. Additionally, we quantify the expected precision of the measurements and how the resulting data would indicate the existence of a dark photon. Finally, we discuss potential challenges of the proposed search at Mu3e.
\end{abstract}

\maketitle

\section{\label{sec:level1} Introduction}

The evolution of the Standard Model throughout the late 20$^{\text{th}}$ century was driven largely by experiments involving increasingly high collision energies, elucidating particle dynamics at higher and higher mass scales. Complementing efforts to probe these dynamics at increasingly high energies, a variety of different experimental setups at lower energies exist to test theories that go beyond its framework. Despite these efforts, the Standard Model has remained remarkably stalwart, with minimal experimental challenges to its consistency in recent years.

In 2016, the Atomki collaboration reported an anomaly in the angular correlation of $e^{+}e^{-}$ pairs that were emitted in the decay of excited beryllium nuclei to their ground states \citep{krasznahorkay2016observation}. This same anomaly was replicated in excited helium nuclei in 2019 by the same group using a revised experimental apparatus \citep{krasznahorkay2019new}. While these results have not yet been replicated, they suggest potential physics beyond the Standard Model. Of the handful of theoretical explanations that suggest new physics, one leading explanation is the presence of dark photon ($A'$): a sub-GeV mass boson that is a candidate carrier for a fifth fundamental force with strong theoretical ties to dark matter \citep{feng2017particle, alves2018viable}. A conclusive dismissal or confirmation of such a particle's existence would either imply an unknown explanation for the Atomki result, such as novel nuclear effects, or lay the foundation for a significant extension of the Standard Model and renewed considerations in cosmology, particle physics, and nuclear interactions.

A a sub-GeV mass boson that couples to leptons would also conveniently align with several other experimental anomalies. The Brookhaven AGS experiment 821 \citep{bennett2006final} measured the magnetic moment of the positive muon to be 3.6$\sigma$ above the value predicted by the Standard Model, and suggests the existence of a $U(1)$ gauge symmetry corresponding to the existence of a dark photon as a potential explanation \citep{carone2013flavor}. A vector boson with dark sector properties has also been proposed as an explanation for the cosmological lithium problem (the discrepancy in predicted and observed abundances of helium and hydrogen isotopes in the universe) as the decay of a dark photon to $e^{+}e^{-}$ pairs would account for the observed deviation \citep{fradette2014cosmological}.

\begin{figure}[H]
  \centering
  \includegraphics[scale=0.29]{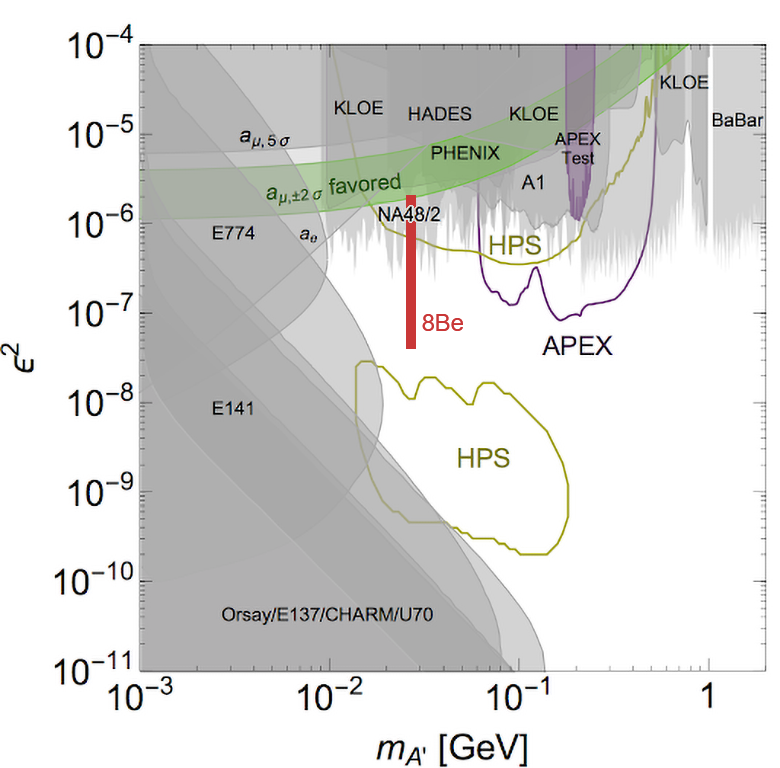}
  \caption{The currently explored $m_{A'}$ - $\epsilon$ parameter range for dark photons. The region in which a dark photon that could explain the Atomki anomaly would live is shown by the red vertical bar \citep{beacham2014apex}.}
  \label{fig:oldExclusion}
\end{figure}

Since the Atomki results from 2016 were published, they have been met with varying degrees of skepticism. The Atomki team reported neither error bars for either experiment nor established limits on the emission rates of new particles, and it is not currently known to what extent systematic errors impact those measurements. The results presented a mass and lepton coupling (a constant that characterizes a particles interaction with leptons) range in which the suspected particle could live. To date, very few experiments beyond the Atomki collaboration have directly probed this parameter range as shown in Figure \ref{fig:oldExclusion}. No experiment that can sweep the entire parameter space of mass and lepton coupling has yet been launched. Additionally, while the discovery of a dark photon would introduce novel physics into the Standard Model, existing experimental results are limited to effects demonstrated in atomic nuclei, which may be confounded by unknown nuclear physics.

In order to eliminate systematic errors and ambiguity that plague prior experiments, we propose an experiment to cross-purpose Mu3e, a particle physics experiment used chiefly to analyze muon decays. Mu3e has the following optimal properties for such a search:

\begin{enumerate}
    \item The capability to search the full extent of mass ($m_{A'}$) vs. lepton coupling ($\epsilon$) parameter space suggested by the anomaly observed by the Atomki result.
    \item Avoids exotic nuclear effects that may have influenced the Atomki result by utilizing muon decays instead of excited nuclei as a source of dark photons.
    \item Provides excellent measurement resolution, which makes it possible to distinguish different signal decays from other background processes to a high degree of accuracy.
\end{enumerate}

 We suggest leveraging Mu3e to perform a novel two pronged search. A resonance search will look to discern the mass signature of a dark photon, while a vertex displacement search will look for dark photons that travel detectable distances before they decay. Unlike the Atomki experiments which rely on nuclear decays to reveal the anomaly, our proposed experiment utilizes particle interactions and decays to probe the direct signature of dark photons. The mass and lepton coupling parameter space of a dark photon suggested by the Atomki observations \citep{krasznahorkay2019new, feng2017particle} are well within the projected resolution of Mu3e. The addition of vertex displacement will extend this parameter range to the boundaries of several existing theoretical limits.

A confirmed discovery of a dark photon in the desired range may provide insight into outstanding experimental incongruities with the Standard Model. Disagreements between predictions and measurements of the dipole moment of the muon and the Lamb shift observed in muonic hydrogen elicit theoretical explanations involving an additional intermediate bosonic decay product. Additional work has suggested that this extension to the Standard Model may explain all three phenomena concurrently.

\section{Background}

\begin{figure}
  \centering
  \includegraphics[scale=0.2]{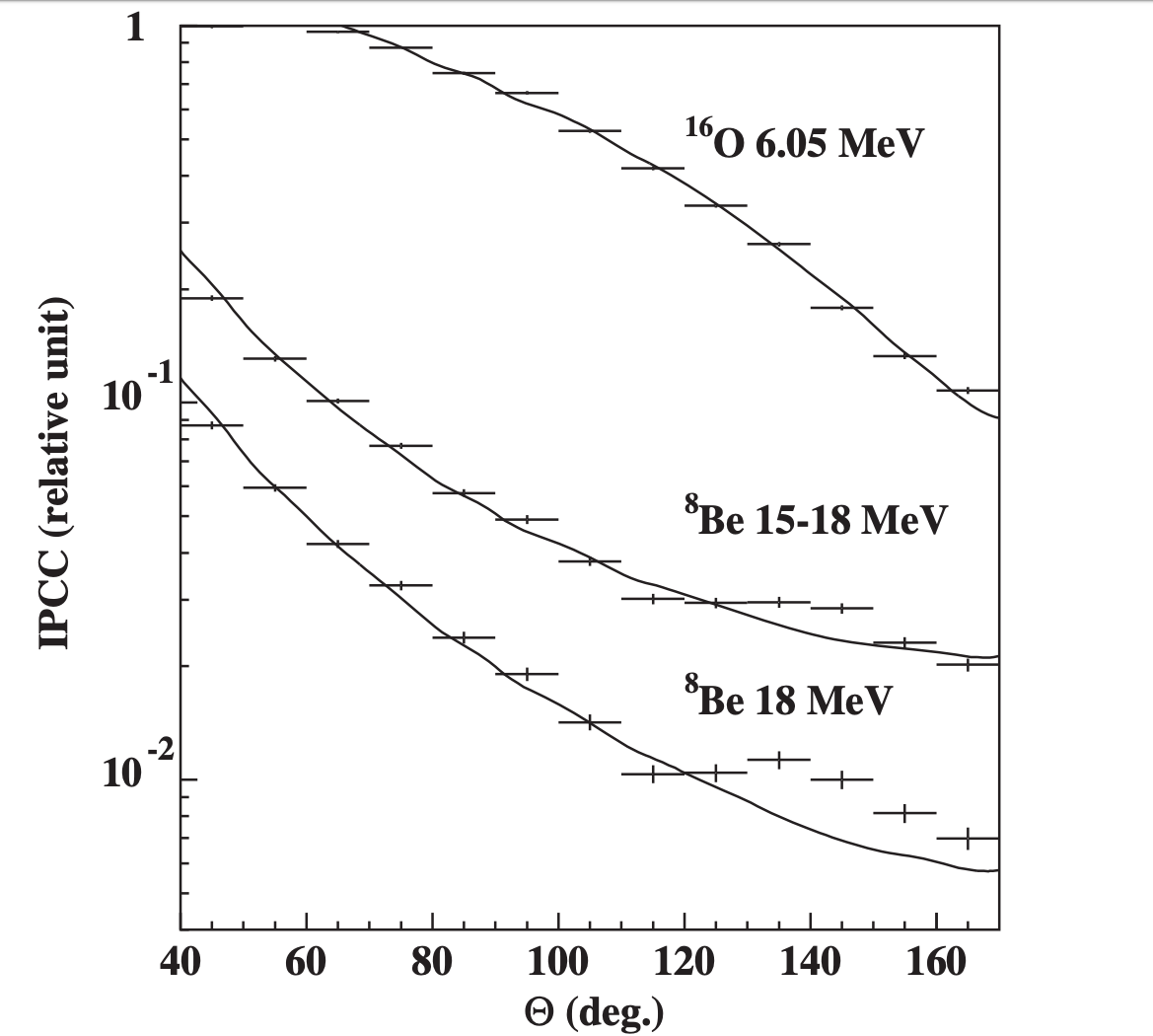}
  \caption{Relative IPCC of simulated Standard Model predictions (solid) versus observed (dotted) measurements versus angular correlation angle. The bump at an opening angle of $\Theta=$140$^{\circ}$ suggests the presence of an intermediate neutral isoscalar decay particle with mass 16.7$\pm$0.85 MeV \citep{krasznahorkay2019new}.}
  \label{fig:atomki_ipcc_li}
\end{figure}

The Atomki beryllium decay experiment \citep{krasznahorkay2016observation} involved the bombardment of a lithium target by a proton beam. Internal pair creation yields $e^{+}e^{-}$ pairs, which are measured by multi-wire proportional counters (MWPCs) in front of each sensor. These readings are used to deduce the spatial correlation \citep{gulyas2016pair} of $e^{+}e^{-}$ pairs. Atomki's initial observations suggested an anomalous bump in the internal pair creation coefficient (IPCC) at a 140$^\circ$ opening angle, visible in Figure \ref{fig:atomki_ipcc_li}.

\begin{figure}
  \centering
  \includegraphics[scale=0.45]{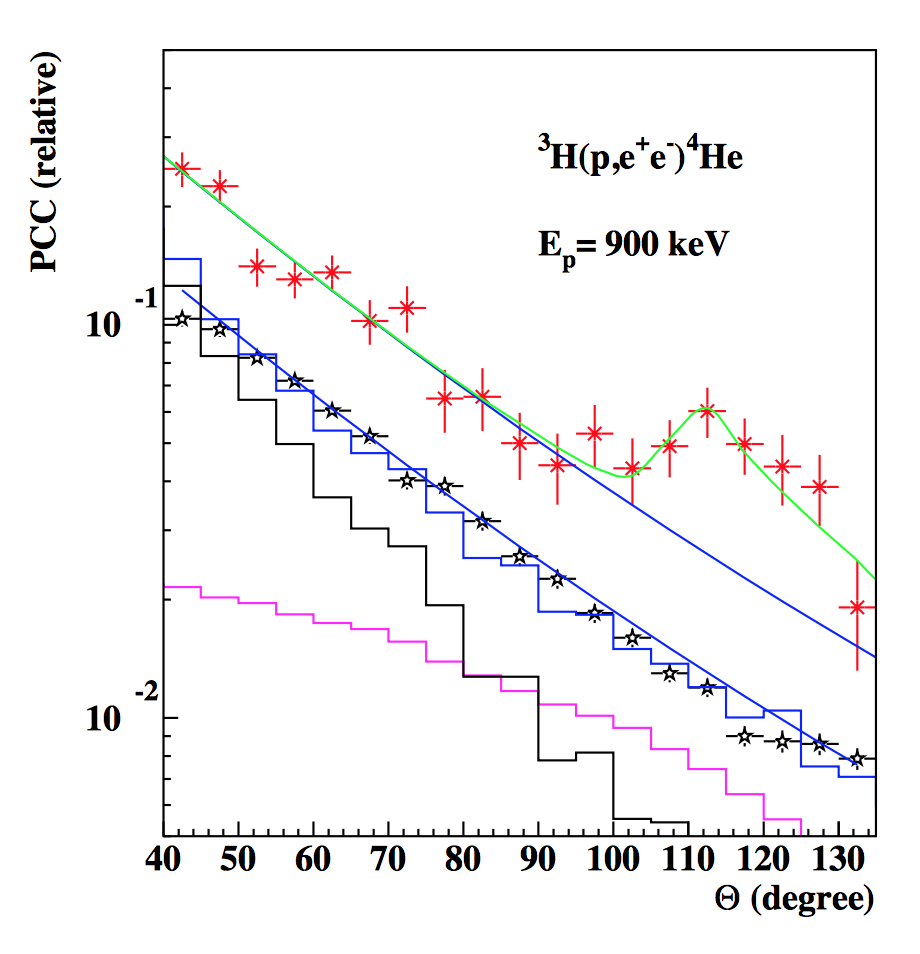}
  \caption{The IPCC correlation versus the opening angle for the $e^+e^-$ pair created in the decay of $^\text{4}$He. The red asterisks show the measured values. An anomalous bump is clearly visible at the $110^\circ$ opening angle for the measured values \citep{krasznahorkay2019new}. This is indicative of an intermediate pseudoscalar decay product of mass 16.84$\pm$0.36 MeV \citep{alves2018viable}.}
  \label{fig:atomki_ipcc_be}
\end{figure}

A second experiment by the same collaboration \citep{krasznahorkay2019new} took place at a separate experimental setup, the Van de Graaff accelerator, and was performed on $^{\text{4}}$He state transitions. This second experiment observed $e^{+}e^{-}$ pairs from the M0 transition of $^{\text{4}}$He, and measured a second peak in IPCC angular correlation indicating an intermediate light decay product of similar mass (see Figure \ref{fig:atomki_ipcc_be}).
This time, the IPCC exhibited an anomalous bump at an angular correlation of $115^\circ$ with a significance of $7.1\sigma$. Despite the high degree of confidence, sources of systematic error in the detector setup or possible alternative explanations for both experiments are not discussed; additionally, background from cosmic decay processes was only shielded up to 50\%. An experiment involving particle interactions eliminates potential confounding effects due to nuclear transitions in $^{\text{8}}$Be or $^{\text{4}}$He.

Follow-up work \citep{feng2017particle} analyzed these results in the context of potential extensions to the Standard Model. Considering the constraints posed by the Atomki observations, theoretical analysis rules out several potential particle explanations and presents a $U(1)$ gauge-symmetric extension to the Standard Model. This extension suggests that the anomaly could be explained by a dark photon detected by $e^{+}e^{-}$ pairs. The properties of such a particle can be summarized in two parameters: its mass $m_{A'}$, and its coupling to leptons ($\epsilon$) which defines the strength of its interaction with Standard Model leptons. Although several dark photon searches have already been conducted, centered at $m_{A'}=16.7$ MeV and $2 \times 10^{-4} < \epsilon < 1.4 \times 10^{-3}$, has not yet been explored in its entirety.

Further experiments are necessary to search specific regions of this mass-coupling parameter space to understand whether a dark photon exists. In addition, any experiment measuring these two parameters must suppress background effects that may produce coincident events. This involves correctly categorizing $A'$ decays and ruling out observed $e^{+}e^{-}$ pairs from different known decay processes.

The NA64 experiment was used to probe a nearby region of parameter space for the existence of a dark photon \citep{banerjee2018search}. This experiment started in 2014 to search for dark photon decays via the decay channel $e^{-}Z\rightarrow e^{-}ZA';$ $A'$ $\rightarrow e^{+}e^{-}$. Throughout 2017 and 2018, no anomalies indicative of a dark photon were found. However, these results excluded the possibility of a 16.7 MeV particle with lepton coupling in the range $1.2 \times 10^{-4} < \epsilon < 6.8 \times 10^{-4}$. Given that earlier theoretical work suggests a dark photon may have $\epsilon$ on the order of $10^{-3}$ \citep{feng2017particle}, the NA64 results narrow down, but do not eliminate the dark photon as an explanation for the Atomki result.

A number of detector facilities are doing ongoing work in nearby areas of parameter space, including LHCb and NA64 at CERN \citep{banerjee2018search}, VEPP-3 at the Budker Institute \citep{wojtsekhowski2018searching}, DarkLight at MIT \citep{katzin2012darklight}, and Mu3e at PSI \citep{echenard2015projections}. Of these experiments, Mu3e is a particularly excellent candidate for a dedicated dark photon search in the area of parameter space associated with the $^{\text{8}}$Be anomaly.

\subsection{Theory}
\subsubsection{Interactions With The Standard Model}

The behavior of a hypothesized dark photon can be expressed by adding the following Lagrangian to the Standard Model Lagrangian:

\begin{align}
    \mathcal{L}_A=-\frac{1}{4}F'^{\mu\nu}F'_{\mu\nu}+\frac{1}{2}m_{A'}^2A^{\mu\nu}A_{\mu\nu}+\epsilon A'^{\mu}J_{\mu}
\end{align}

The first two terms correspond to the kinetic and mass terms, respectively, and the third term describes lepton interactions. A dark photon can then be characterized by its mass ($m_{A'}$) and lepton coupling ($\epsilon$) \citep{fradette2014cosmological}.

\subsubsection{Dark Photon-Mediated Pair Creation}

\begin{figure}[ht]
  \centering
  \includegraphics[scale=0.5]{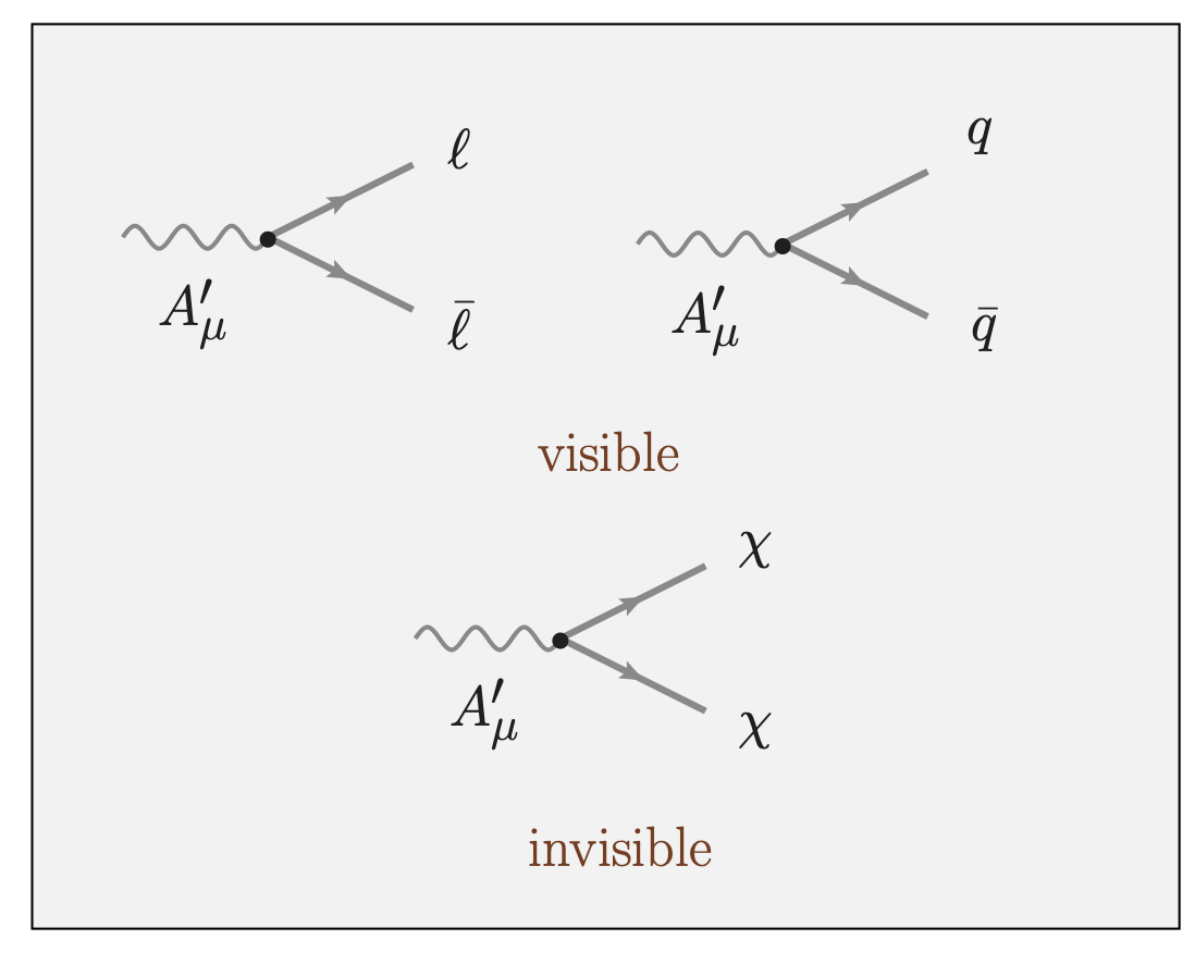}
  \caption{Massive dark photons can take three different decay channels. In the visible sector it can decay into either lepton-antilepton or hadron-antihadron pairs. It may also decay into other hypothetical dark sector particles undetectable by standard methods. \citep{fabbrichesi2020dark}}
  \label{fig:decay_channels}
\end{figure}

Massive dark photons can decay into other particles via three hypothetical channels (Figure \ref{fig:decay_channels}). The two visible decay channels consist of a dark photon decaying into a lepton-anti lepton pair or a hadron-anti hadron pair. Furthermore, dark photons can decay into other hypothetical dark sector particles through what is known as an `invisible decay,' as none of the particles in the decay process can be directly detected through successive interaction with known Standard Model particles. \citep{fabbrichesi2020dark}.

Our proposed experiment aims to search for dark photons in the mass range $10\text{ MeV}\leq m_{A'}\leq 80\text{ MeV}$. This mass range is chosen specifically since it aligns with most theoretical predictions, including the results of the Atomki experiment. Since pions ($\pi$) are the lightest hadron, hadron decay is prohibited for $m_{A'}<m_{\pi}$ by energy conservation, and all visible $A'$ decays must follow the lepton decay channel. By the same argument, decay into particle-antiparticle pairs of other lepton families such as $\tau\bar{\tau}$ (2$\cdot$1777 MeV) and $\mu\bar{\mu}$ (2$\cdot$105.7 MeV) pairs is also prohibited, since their combined invariant masses are significantly higher those of the probed $A'$ mass range. This leaves $A'\rightarrow e^{+}e^{-}$ as the only viable candidate for decay into Standard Model particles. The decay width for this channel is given by:

\begin{align}
    \Gamma_{A'\rightarrow e^{+}e^{-}}=\frac{\alpha\epsilon^2}{3}m_{A'}\sqrt{1-\frac{4m_e^2}{m_{A'}^2}}\left( 1+\frac{2m_e^2}{m_{A'}^2}\right)
\end{align}

An experiment can then detect the production of dark photons by measuring the output of charged electron and positron decay products, along with the change in 4-momentum during any decay processes. The momentum of a $e^{+}e^{-}$ pair traveling in different directions can be related to the pair's invariant mass in the lab frame as:

\begin{align}
    m_{e^+e^-}^2&=(E_{e^+}+E_{e^-})^2-(\overline{p}_{e^+}+\overline{p}_{e^-})^2
\end{align}

Assuming that these electrons decayed from a stationary $A'$ (in the lab frame of reference) the invariant mass would simply be $m_{A'}$ which can be related to the emission angle and electron energies as:

\begin{align}
\begin{split}
    m_{A'}&=m_{e^+e^-}\\
    m_{A'}^2&=\hspace{0.1cm}m_{e^+}^2+m_{e^-}^2+2E_{e^+}E_{e-}\\
    &+2\sqrt{E^2_{e^-}-m_{e^-}^2}\sqrt{E^2_{e^+}-m_{e^+}^2} cos(\theta)
\end{split}
\end{align}

A plausible dark photon signature would then be the emission of $e^{+}e^{-}$ pairs, with a total invariant mass equal to that of the dark photon.

\subsection{Interactions between Muons and Dark Photons}

Dark photons can serve as intermediary particles in
any reaction where a virtual photon would otherwise be emitted, if the decaying particle carries sufficient energy. In the
context of muon decay, this provides a candidate decay mode with an intermediate dark photon decay product: radiative muon decay (RMD).

In Standard Model physics, RMD takes the form $\mu^+\rightarrow e^+\nu_e\overline{\nu_\mu}\gamma$. An output photon $\gamma$ of sufficient energy can then decay further to a $e^{+}e^{-}$ pair: $\gamma\rightarrow e^+e^-$. Theoretically, a dark photon could take the place of the Standard Model photon in this same process $\mu^+\rightarrow e^+\nu_e\overline{\nu_\mu}A'$, $\text{ }A'\rightarrow e^+e^-$.

The decay can take one of three possible pathways (see Figure \ref{fig:dark_photon_feynman}) with the dark photon decaying from the muon, the positron or an intermediate $W$ boson. The last process is suppressed by the low mass ratio of the muon to the $W$ boson, on the order of $~10^{-6}$. \citep{echenard2015projections}. However, the first two channels can be observed experimentally and have a branching ratio of:

\begin{align}
    B_{\text{lep}}&=\frac{1}{3\times 10^{-19}}\left(\frac{\epsilon}{0.1}\right)^2\text{exp}\left(\sum_{i=0}^5 a_i\left(\frac{m_{A'}}{\text{GeV}}\right)\right)
    \label{ap_branching_ratio}
\end{align}

where $\{a_i\}_{i=0}^5$ are constants presented in \citep{echenard2015projections}.

Equation (\ref{ap_branching_ratio}) indicates that the decay process of the $A'$ through leptons has a fixed decay width that depends solely on the electron mass $m_e$, $m_{A'}$ and the lepton coupling $\epsilon$. Using this decay width, we can then estimate the time constant $\tau_{A'\rightarrow e^{+}e^{-}}$ and mean decay length via Equation (\ref{ap_time_constant}), assuming a propagation speed of $c$.

\begin{align}
    c\tau_{A'\xrightarrow{}e^+e^-}&\approx0.8\text{mm}\left(\frac{10^{-4}}{\epsilon}\right)^2 \frac{10\text{ MeV}}{m_{A'}}
    \label{ap_time_constant}
\end{align}

\begin{figure}
  \centering
  \includegraphics[scale=0.25]{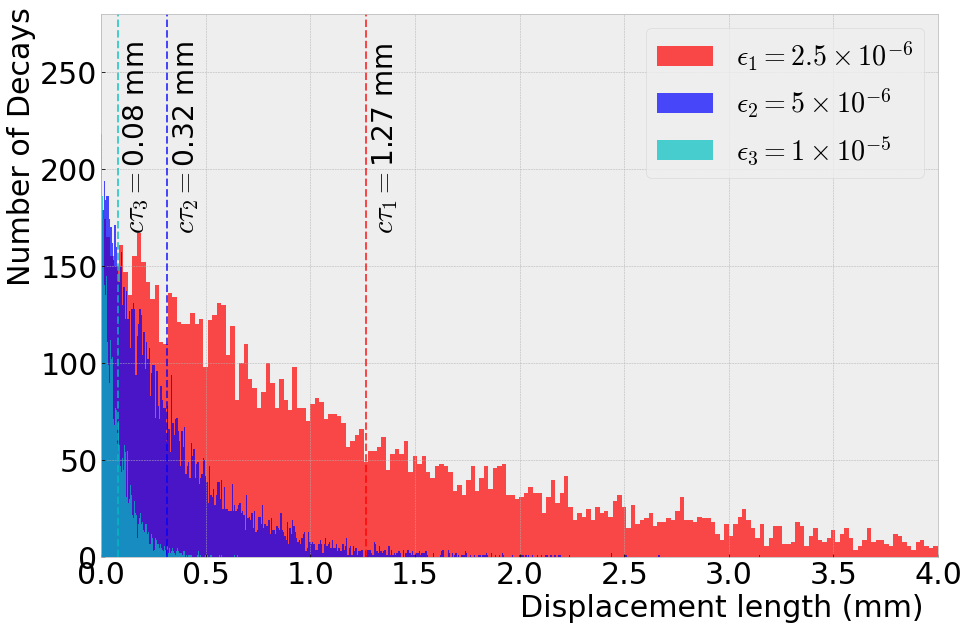}
  \caption{Simulation of $A'$ decay lengths for an $A'$ with $m_{A'}=10\text{ MeV}$ and three different values of $\epsilon$. There is a clear increase in mean decay length with a decrease in kinetic coupling.}
  \label{fig:decay_number_simulation}
\end{figure}

The above relation indicates that for smaller values of $\epsilon$, we would expect to see larger decay lengths. Given adequate vertex resolution and sufficiently low $\epsilon$, the vertex of the dark photon at decay would be distinguishable from the decay vertex of the muon itself. This makes it possible to recognize dark photon decay processes by the presence of an additional decay vertex. The simulation in Figure \ref{fig:decay_number_simulation} depicts the simulated distribution of $A'$ decay lengths for $m_{A'}=10\text{ MeV}$ and three different coupling constants: $\epsilon_1=2.5\times10^{-6},\epsilon_2=5\times10^{-6}$, and $\epsilon_3=1\times10^{-5}$. The respective mean decay lengths for each coupling constant are $1.27\text{ mm}, 0.32\text{ mm}$, and $0.08\text{ mm}$. This indicates that there is a clear increase in the mean decay length for smaller values of $\epsilon$, and enables a dark photon search at smaller $\epsilon$ value than those perceptible in a standard mass bump hunt, which we discuss in the following section.

\subsection{Searching for the Dark Photon}

\subsubsection{Resonance Search}

Given the relationship between the invariant mass of the emitted $e^{+}e^{-}$ pair and the mass of the dark photon, we can perform a `bump hunt' for RMDs mediated by a hypothetical $A'$ particle. Standard Model muon decay processes including Michel decay ($\mu^{+}\rightarrow e^{-}\nu_{e}\bar{\nu}_e$) and Bhabha and Compton scattering also release electrons and positrons with high branching fractions and tend to dominate the invariant mass spectrum of muon decay experiments.

\begin{figure}[ht]
  \centering
  \includegraphics[scale=0.5]{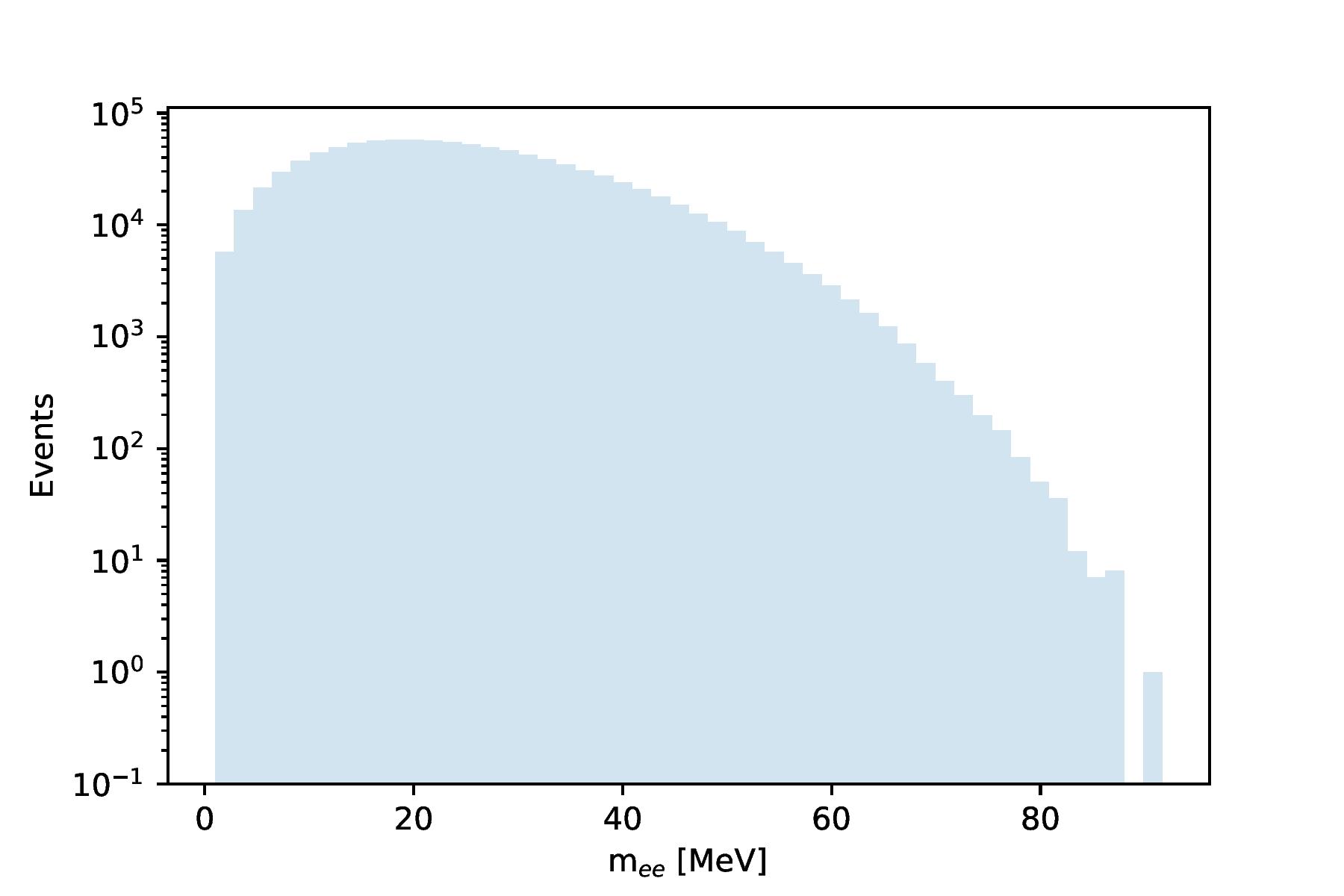}
  \caption{Invariant mass distribution of decay process $A'\rightarrow e^{+}e^{+}e^{-}\nu_e\bar{\nu}_\mu$ for a simulated dark photon decay process. There is a clear peak in the invariant mass spectrum at $\sim 17\text{ MeV}$, the signature of $e^{+}e^{-}$ pairs emitted by a dark photon.}
  \label{fig:signal_spectrum}
\end{figure}

As discussed in Section II(A), the indication of a dark photon decay would be the constant invariant mass of emitted $e^{+}e^{-}$ pairs. This would appear as a peak at a particular mass over the background invariant mass spectrum. We simulated $10^6$ decays of a dark photon ($m_{A'}=17\text{ MeV}$, $\epsilon=10^{-4}$) into $e^{+}e^{-}$ pairs and graphed the resulting invariant mass spectrum in Figure \ref{fig:signal_spectrum}. As expected there is a strong peak in the invariant mass at the mass of the mediating dark photon. 
The primary challenge with identifying this peaked invariant mass signal is distinguishing it from the invariant mass spectrum of other decay processes that may occur within the detector. We will address this challenge in Section III.

\subsubsection{Vertex Displacement Search}

\begin{figure}
  \centering
  \includegraphics[scale=0.25]{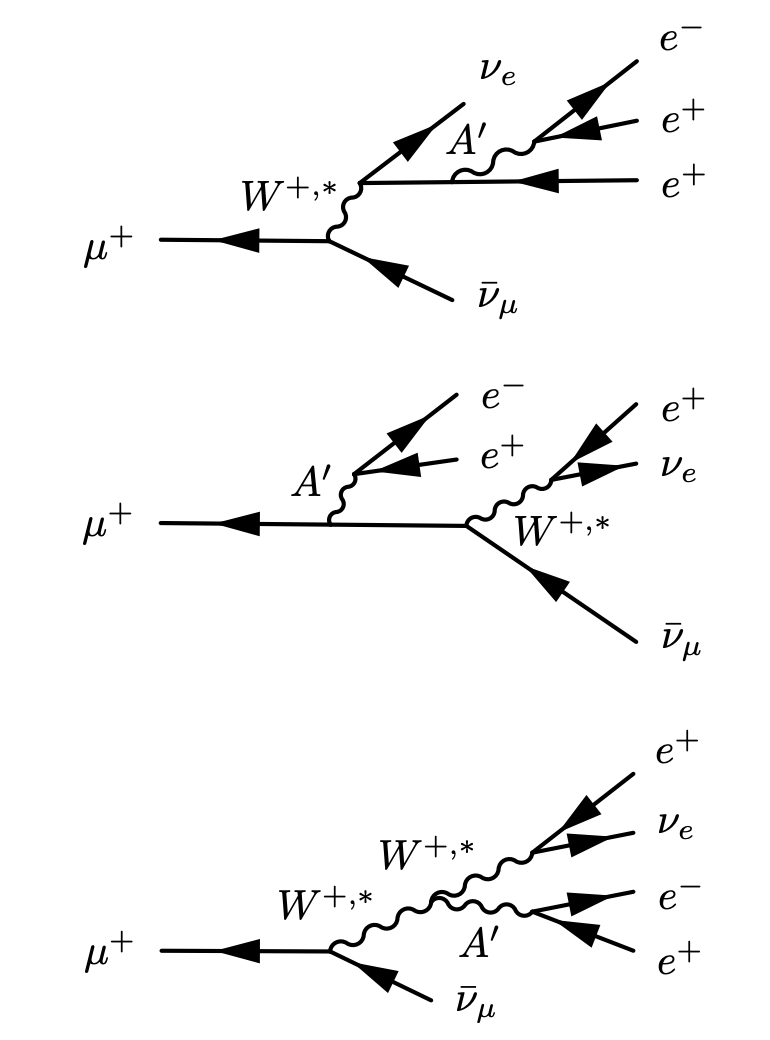}
  \caption{Feynman diagrams for the three conjectured on-shell muon decay processes that decay via a dark photon. The third process is suppressed to the order of $m_\mu^2/m_W^2 \sim 10^{-6}$ \citep{echenard2015projections}}
  \label{fig:dark_photon_feynman}
\end{figure}

In Standard Model muon decay processes, the output photon is uncharged and stable, and so undetectable to a charged particle sensor. On the other hand, an $A'$ decay product is unstable and will decay via the decay channel $\mu^{+}\rightarrow e
^{+}\nu_e\bar{\nu}_\mu A'; A'\rightarrow e^{-}e^{+}$ \citep{echenard2015projections}, where the corresponding Feynman diagrams are pictured in Figure \ref{fig:dark_photon_feynman}.

If the dark photon's lepton coupling is sufficiently small, the location of its decay will be measurably displaced from the stopping location of the muon. This displacement can be used to deduce the dark photon lifetime, which can be used to solve for the lepton coupling via Equation (\ref{ap_time_constant}). The invariant mass of the dark photon can also be deduced by summing the 4-momenta of its decay products. In this way, both the invariant mass and lepton coupling of a dark photon can be deduced by vertex reconstruction.

However, several additional factors further complicate this. Firstly, background processes can yield virtually identical signals and must be suppressed to correctly parse out the dark photon signal. Standard Model photons emitted by RMD ($\mu^{+} \rightarrow e^{+} \gamma \nu_e \bar{\nu_\mu}$) can scatter off atomic nuclei and result in $e^{+}e^{-}$ pairs like $A'$-mediated dark photon decay. This scattering is essentially random, and will not show the same peaked invariant mass signature of $e^{+}e^{-}$ pairs of a dark photon. Furthermore, the displacement of scattered vertices will be roughly uniformly distributed in the vicinity of the target, in contrast to the anticipated exponential decay associated with the Poisson process of an $A'$ particle.

Secondly, for smaller values of $\epsilon$ the branching factor for electron-mediated dark photon decay decreases proportionally to $\epsilon^2$. As a result, smaller values of $\epsilon$ elicit fewer decay events. However, the proportionately increasing decay length $c\tau$ makes it possible to reject background signals more accurately in the form of narrow cuts on the data. Consequently, despite a smaller signal, we can expect a better signal-to-noise ratio for smaller values of $\epsilon$.

Finally, distinguishing the source of the positrons from the $A'$ and other background processes can potentially be severely challenging as there is no universal method by which to discriminate them by process. Knowing which interaction generated a positron is crucial, as this is the basis identifying the separate decay vertices using measurements of output products. However, a combinatorial analysis of different reconstructions can indicate the presence of an invariant mass spike, which can be used to distinguish which pair might have originated from the decaying dark photon. Furthermore, analysis of $W$ boson decay at CERN has shown that positrons emitted in this process have a well-characterized momentum distribution with a large majority of ejected positrons being emitted nearly perpendicular to the propagation of the W boson with high energy \citep{watkins1986discovery}. A highly collimated muon beam will also provide strong bounds on the transverse distribution of stopped muons, which enhances the signal-to-noise ratio of any externally distributed vertices. These features provide a basis for probablistically estimating the likelihood of a measured vertex displacement corresponding to an anomalous dark photon.

\section{Description Of The Research Instrument}

\subsection{Overview of Mu3e Experiment}

Mu3e was initially conceived to search for the violation of lepton flavor conservation, specifically by testing for the decay process $\mu^{+}\rightarrow e^{+}e^{-}e^{+}$. A novel tracking concept, in conjunction with high momentum and spatial resolution, equips the detector with the ability to efficiently search for this process through a similar resonance search. Lepton flavor violation in muon decay has a theorized branching fraction of $~10^{-16}$ which Mu3e is capable of discerning \citep{berger2014mu3e}. This extreme sensitivity makes it an ideal candidate for our experiment.

The detector consists of a narrow, high-intensity muon beam impinging on an aluminium target in the shape of a hollow double cone. This target will stop an estimated $\sim$ $10^8$ muons per second in the first phase of the experiment. The entire detector is enveloped in a 1 T solenoidal magnetic field parallel to the beam direction. Stopped muons yield charged decay products, which are accelerated in helical paths by the magnetic field and picked up by the tracking system.

\begin{figure}
  \centering
  \includegraphics[scale=0.3]{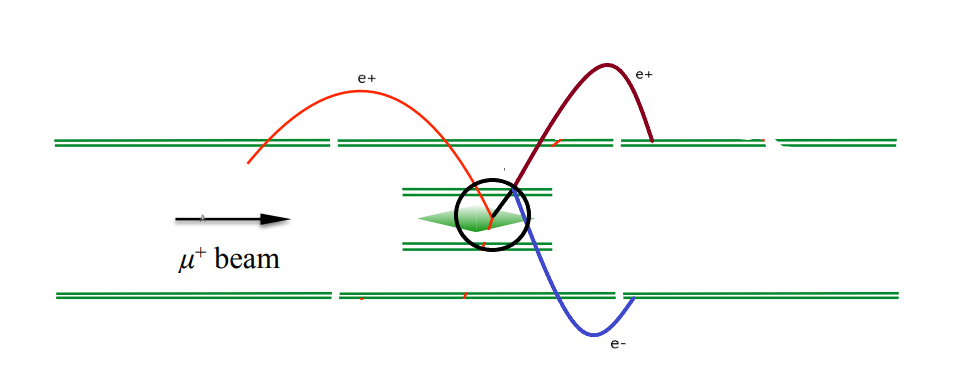}
  \caption{Schematic of muonic dark photon decay in the Mu3e detector, with the two graphite targets in the center. A muon stopped inside the target may theoretically emit an $A'$ with mean decay radius (black circle) and a positron (red). This $A'$ then decays with some probability into an $e^{+}e^{-}$ (blue/maroon) pair. The solenoidal magnetic field in the detector causes the electrons to travel in helical paths through the twin layers of pixel and tile sensors (green), with which their momenta can be measured.}
  \label{fig:mu3e_target_diagram}
\end{figure}

The tracking system comprises of two detection layers, each composed of MUPIX pixel sensors with scintillating tiles sandwiched in each layer (see Figure \ref{fig:mu3e_target_diagram}). These custom-designed pixel sensors measure the location and charge of electrons and positrons that travel through the detector with an axial resolution of $\sigma_{x,y,z}\approx30\mu\text{m}$ while the scintillating tiles measure the timestamp of particle hits with a time resolution of $O(10\text{ ps})$.

\subsection{Motivation for Mu3e}

The primary motivations for the utilization of Mu3e as a candidate detector are twofold. Firstly, the intensity of the pion beam generated from the $\pi e5$ project at PSI provides a very intense source of muons which affords Mu3e a vast muon sample size, allowing the experiment to probe further decays with extremely small branching factors. Secondly, the state-of-the-art tracking and timing sensors built into the detector, including the MUPIX silicon tracking sensors and scintillating fibre/tile timing systems, provide excellent spatial and momentum resolution. This is an integral requirement of both vertex reconstruction and bump hunt approaches. This makes it possible to distinguish events to the degree necessary to test for the occurrence of interactions with an intermediate $A'$, to a high degree of confidence.

\subsubsection{Statistics}

\begin{center}
    \textbf{\textit{Phase I}}
\end{center}
The Phase I iteration of the Mu3e detector uses muons sourced from pion decays in the $\pi e5$ project. We anticipate an available muon intensity between $0.7-1.0\times10^8 \text{ muons/s}$ \citep{berger2014mu3e} which accounts for $1.18-2.59\times10^{15}$ total stationary muons over the course of the proposed 300-day runtime. Using the branching factor for non-mass suppressed $A'$-mediated RMDs (Equation \ref{ap_branching_ratio}), we can calculate bounds on the event rates for a given $m_{A'}$ and $\epsilon$.

For the coupling and mass values of $15\leq m_{A'}\leq20 \text{ MeV}$ and $2\times10^{-4}<\epsilon<9\times10^{-4}$ suggested by the Atomki anomaly, this would translate to branching fractions from $2.14\times10^{-9}$ to $7.24\times10^{-8}$. This translates to between $5.5\times10^6$ and $1.8\times10^8$ $A'$-mediated muon decay events over the course of Phase I. Based on the simulation in Figure (\ref{fig:decay_number_simulation}), this will be sufficient to distinguish a potential dark photon signal from accidental and background processes with statistical hypothesis testing.

\begin{center}
    \textbf{\textit{Phase II}}
\end{center}

The Phase II iteration of Mu3e hopes to increase the number of muons generated through integration with the High Intensity Muon Beam Project \citep{blondel2013research}. Pending the completion of this project, Mu3e expects to source a significantly higher muon current, yielding a muon stopping rate of approximately $3\times10^{10}$ muons/s. Once again using the $A'$ mass and kinetic coupling values motivated by the Atomki experiment, we anticipate between $1.6\times10^9$ and $5.5\times10^{10}$ total $A'$-mediated muon decay events.

\subsubsection{Measurement Resolution}

The nature of the proposed two-pronged search with a combined resonance and vertex displacement search requires excellent momentum space and vertex resolution. This section will discuss the current measurement abilities of Mu3e and elucidate the advantages and limits that it provides us.

\begin{center}
    \textit{\textbf{Momentum Resolution}}
\end{center}

Having been constructed specifically to distinguish lepton flavor-violating muon decay, the Mu3e detector provides ideal momentum resolution for distinguishing this decay mode from background processes. Our resonance search also depends on the same high momentum resolution to distinguish $A'$-mediated RMD from other possible decay modes. Given the significantly higher branching ratio of predicted $A'$-mediated muon decay events relative to lepton flavor-violating muon decay, plus the fact that the Mu3e detector as it stands maintains a $0.5\text{ MeV}$ momentum resolution, the Mu3e detector can reliably distinguish between the $A'$-mediated events and Standard Model events (see Section III (c)).

\begin{center}
    \textbf{\textit{Vertexing Resolution}}
\end{center}

\begin{figure}
  \centering
  \includegraphics[scale=0.25]{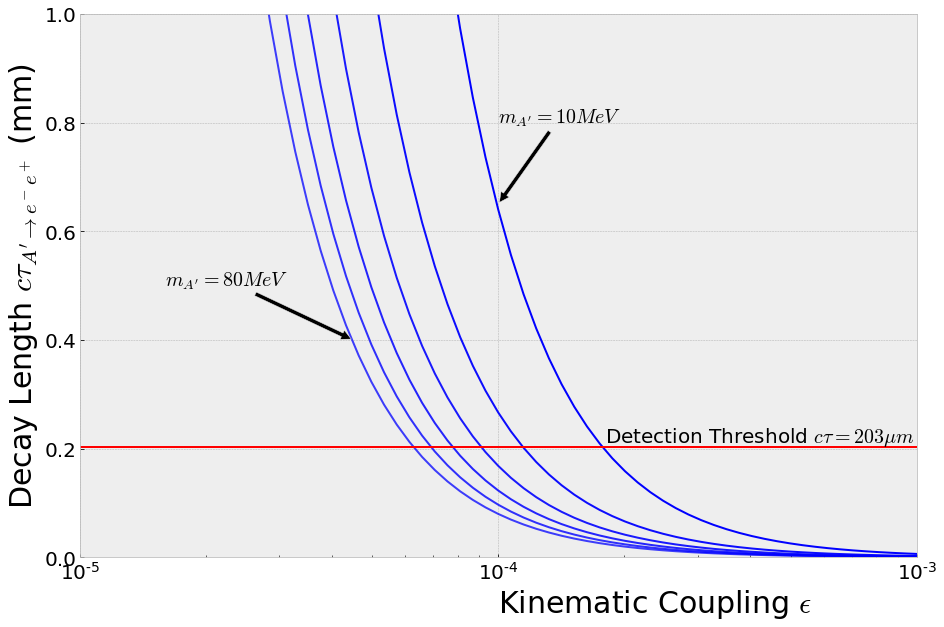}
  \caption{Plot of the mean decay length versus lepton coupling ($\epsilon$) for dark photon masses between $10$ MeV and $80$ Mev in $14$ MeV increments. Decay lengths above the detection threshold can be clearly identified with the current vertex resolution at Mu3e. This highlights that for coupling values below $\epsilon=7\times10^{-5}$ we will be able to detect a displaced vertex for any mass in our parameter range.}
  \label{fig:threshold}
\end{figure}

Another aspect of the existing detector that complements our proposal is its high vertexing resolution. The current detector setup expresses a vertex resolution of $\sigma_{x,y}=230\mu\text{m}$ and $\sigma_z=320\mu\text{m}$ \citep{augustin2019mupix8}. Using the Shapiro-Wilks unimodality metric along with a significance limit of $5\sigma$, we see that any unimodality in the data can safely be ignored at average decay lengths above $203\mu\text{m}$, which we can label as our detection threshold. With a mean decay length described by Equation (\ref{ap_time_constant}), we construct a graph of possible mass-coupling combinations accessible to a vertex displacement search (see Figure \ref{fig:threshold}).

Figure \ref{fig:threshold} conveniently provides an upper bound on what $\epsilon$ range a vertex displacement search could explore. The lower bound on $\epsilon$ would be determined by the interplay between the increasing mean decay length and the falling branching factor. The Mu3e geometry requires that any charged particles pass through at least three pixel sensors (inner layer once and the outer layer twice) to be able to measure momentum. If the decay length exceeds the radius of the inner layer we will no longer have three pixel sensor hits making it impossible to reconstruct these vertices. Given an inner layer radius of 2 cm, for $\epsilon>6\times10^{-10}$ the mean decay length roughly equals the radius. However at values of $\epsilon$ of this scale, the branching factor is $\sim10^{-19}$.
With Mu3e designed to probe events with branching factors on the order of $>10^{-16}$, dark photons in this $\epsilon$ scale would not be detectable. A conservative lower bound of $\epsilon\approx5\times10^{-7}$, will provide us with a mean decay length of $\sim2\text{ cm}$ and a branching factor of $\sim10^{-13}$, which are safely within Mu3e's capabilities to identify and reconstruct \citep{berger2014mu3e}.

Another advantage of Mu3e is the prospect of using a highly collimated muon beam provided by $\pi e5$. With a nominal standard deviation in intensity of $~200\mu\text{m}$ in the transverse plane and strong unimodality produced by using a mixture of beam stops and collimators, we can expect a large majority of muon decays to occur in a very concentrated region in the transverse plane. Thus a vast majority of muons will be stopped within the detection threshold of $203\mu m$. This in turn means that signal events detected beyond the threshold are much more likely to be true displaced $A'$ vertices, rather than promptly decaying $A'$ vertices formed by muons stopped outside the detection threshold. 

\subsection{Experimental Backgrounds}

\subsubsection{Background Decay Processes}

In addition to the signal decay process of $A'$-mediated RMD $\mu^{+}\rightarrow e^{+}\nu_e\bar{\nu}_{\mu}A',A'\rightarrow e^{-}e^{+}$, there are several competing background processes that yield similar decay products. Given the $\sim$29 MeV energy of incident muons in the detector, processes that also generate $e^{+}e^{-}$ pairs in the Standard Model include RMD, Michel decay, and Bhabha scattering. In this section we discuss these decays, and how we propose to suppress them and extract a clear signal-to-noise ratio.

\begin{center}
    \textit{\textbf{Michel Decay}}
\end{center}

\begin{figure}[H]
  \centering
  \includegraphics[scale=0.5]{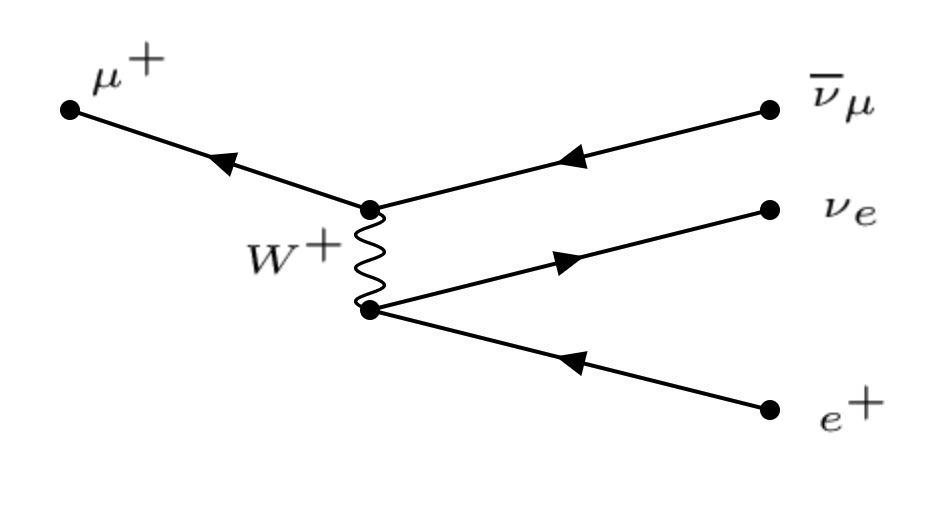}
  \caption{Feynman Diagram of Michel decay. Given the polarized input muons, this process would only yield positively charged positrons, making it straightforward to isolate from the signal decay.}
  \label{fig:michel_feynman}
\end{figure}

Michel decay ($\mu^{+}\rightarrow e^{+}\nu_e\bar{\nu_\mu}$) is the primary muon decay mode, with a branching factor of virtually 1 \citep{berger2014mu3e}. Decay products include a positron, electron antineutrino, and muon neutrino. With only a single charged particle emitted, Michel decay is clearly distinct from the signal decay since a dark photon event would see three charged particles (two positrons and one electron). Given a time resolution of $\sigma_t\approx10^{-10}$ s and a muon stopping rate of $~10^{8}$ muons/s, the majority of these events will be temporally separable and can be rejected on an event-by-event basis.
\newpage
\begin{center}
    \textit{\textbf{Standard Model RMD}}
\end{center}

\begin{figure}[H]
  \centering
  \includegraphics[scale=0.4]{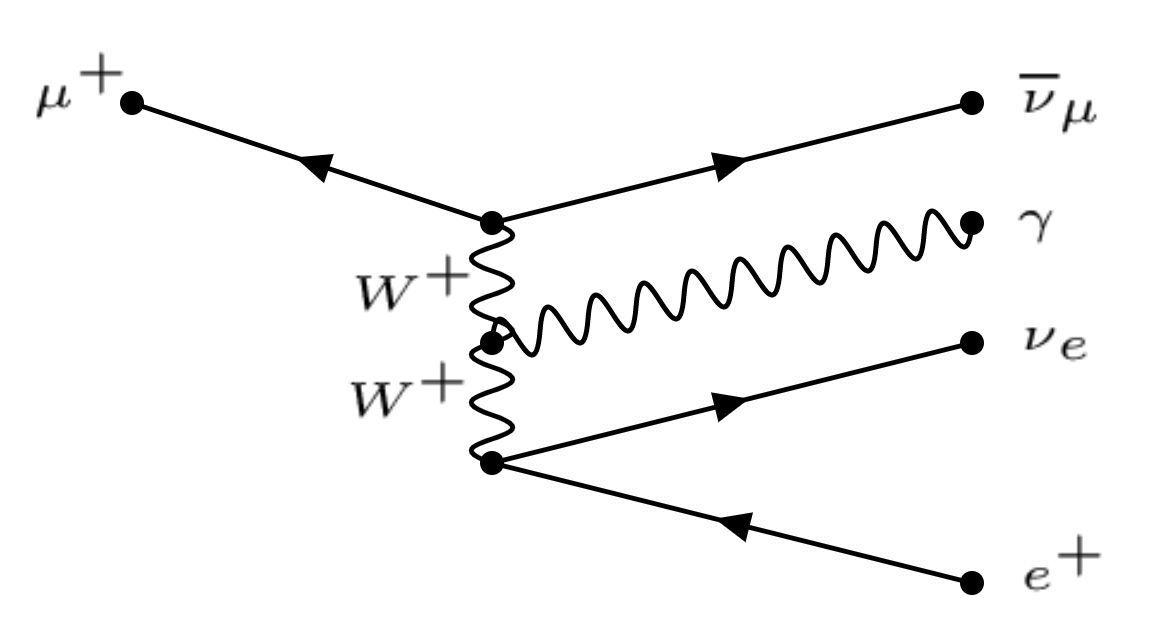}
  \caption{Feynman diagram of RMD. The photon output by this process can decay to a $e^{+}e^{-}$ pair as in $A'$ decays, but would have a distinct invariant mass peak at $m_{e^+e^-}\sim0$ due to the photons lack of a rest mass}
  \label{fig:feynman_rmd}
\end{figure}

As discussed in Section II (B), RMD in the Standard Model is the visible sector equivalent of our signal process with each photon replaced by an $A'$. The identical decay products mean this process cannot be distinguished on an event-by-event basis. Instead, the key distinguishing factor between the two processes is that an $A'$ would have rest mass, in contrast to a visible photon. The invariant mass of $e^{+}e^{-}$ pairs emitted by this decay would then sum to zero. With a branching factor of $~10^{-5}$ \citep{kuno2001muon}, we anticipate that this decay mode will form the majority of observed background. but can be reliably distinguished through invariant mass analysis. Figure \ref{fig:backgrounds} depicts the expected invariant mass spectrum for $5.5\times10^{16}$ RMD events. As expected we see a clear peak at $m_{e^+e^-}=0$.

\begin{center}
    \textbf{\textit{Bhabha Scattering}}
\end{center}

\begin{figure}[H]
  \centering
  \includegraphics[scale=0.3]{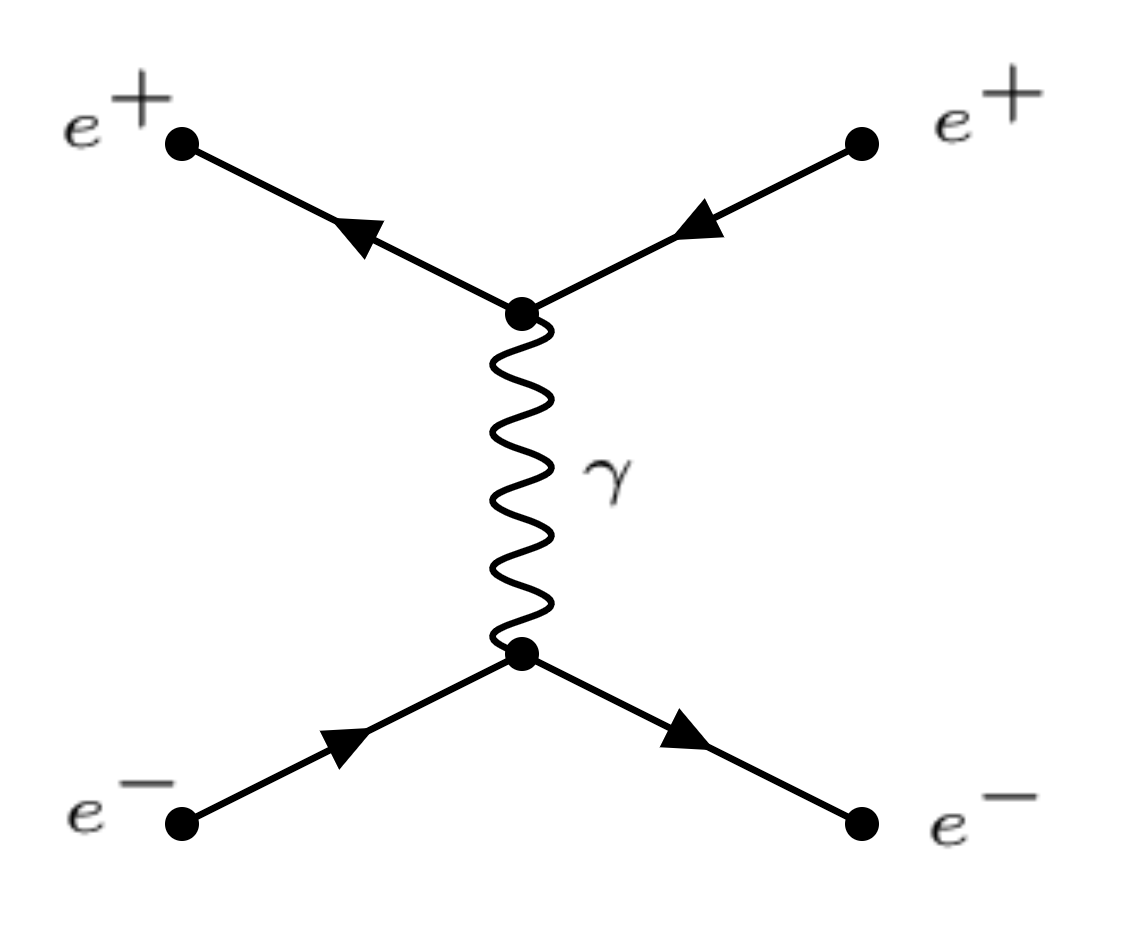}
  \caption{Feynman diagram of Bhabha scattering. This process involves the exchange of a photon between a $e^{+}e^{-}$ pair, which can result in vertex reconstruction yielding a false common vertex for the pair. This process also lacks the invariant mass signature of an $A'$ decay due to the uniform distribution of pair trajectories.}.
  \label{fig:feynman_bhabha}
\end{figure}

Bhabha scattering is a non-muonic background which involves the exchange of energy between a positron and electron through photon emission and reabsorption. Vertex reconstruction for the emitted pair will lead to a common vertex, much in the same way that an $A'$ will emit a positron and electron from a common vertex. However, the invariant mass spectrum will be uniformly distributed due to the lack of correlation between the particle pair. This is unlike the expected peaked invariant mass spectrum of $A'$ decays, allowing us to distinguish this background.

\subsubsection{Accidental Backgrounds}

As established in the previous section, the anticipated background processes can be reliably distinguished from signal decays as isolated processes. However, combinations of these events combined with the imperfect efficiency of the detector can result in events that are impossible to distinguish from true $A'$ decays. For example, three coincident Michel decays coupled with the misidentification of an electron as a positron will be indistinguishable from a signal decay. Estimating the likelihoods of such events is crucial in order to establish the statistical significance of any observed signals. These `accidental' backgrounds are difficult to estimate accurately and even more difficult to verify due to their probabilistic nature. Here we defer to the estimates provided by \citep{echenard2015projections}. This work identifies three main backgrounds that appear identical to a signal decay. Namely:
\begin{enumerate}
    \item Three coincident Michel decays with an electron misidentified as a positron.
    \item Coincident Michel decay and RMDs where the photon decays into a $e^{+}e^{-}$ pair and a positron remains undetected.
    \item Coincident Michel decay and RMD where the photon interacts with the detector bulk to release a $e^{+}e^{-}$ pair via pair creation. One positron remains undetected.
\end{enumerate}

\begin{figure}[ht]
  \centering
  \includegraphics[scale=0.4]{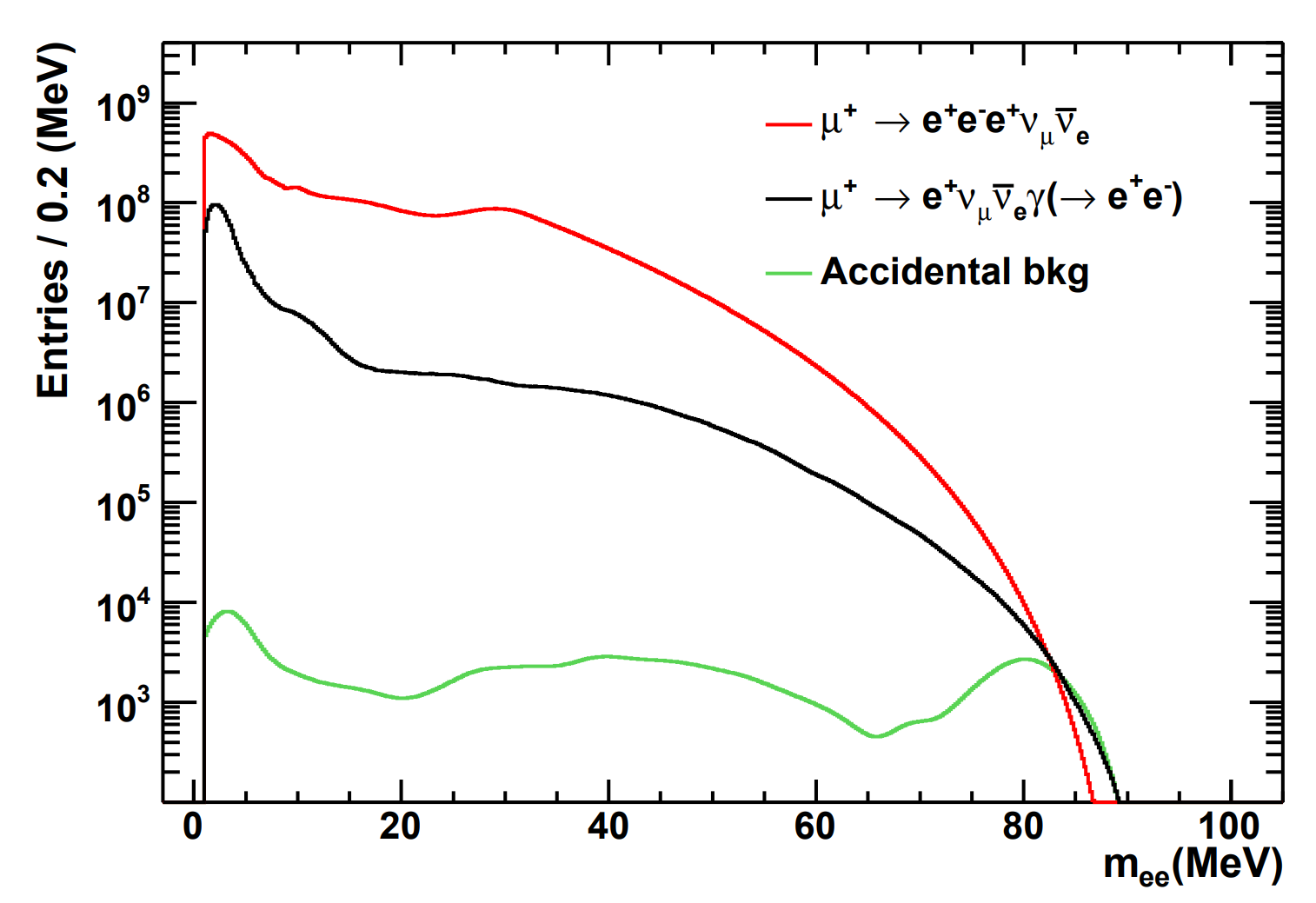}
  \caption{Invariant mass spectrum for the Standard Model muon decay background and accidental background for a simulated $5.5\times10^{16}$ events \citep{echenard2015projections}. As expected, Michel decay and RMD events present an invariant mass that is peaked at $0\text{ MeV}$, unlike a dark photon signal. The accidental background presents a multi-nodal distribution clearly distinct from an anticipated $A'$ signal.}
  \label{fig:backgrounds}
\end{figure}

As highlighted in Figure \ref{fig:backgrounds}, the accidental background presents a multi-nodal distribution that lacks the well-defined invariant mass peak of an $A'$ signal (see Figure \ref{fig:signal_spectrum}). While it is difficult to estimate how well this simulation corresponds to real accidental rates, we can expect a similar skewed invariant mass spectrum. We can further conclude that the accidental background would be several orders of magnitude lower than the Standard Model background, and will consequently be a subdominant source of noise.

\subsection{Risk Management}
The success of our proposed experiment faces three main risks:

\begin{enumerate}
    \item Background processes may have a significant effect on the statistical visibility of signal events.
    \item Unquantified detector effects interfering with our ability to accurately measure decay products and momenta.
    \item Computational feasibility of invariant mass reconstruction and vertex displacement reconstruction.
\end{enumerate}

\noindent As described in previous sections, we are confident that for a vast majority of our targeted parameter space we will be able to mitigate the effects of background events. Further analysis will be required to determine the fine structure of the parameter space we intend to explore and to establish exclusionary confidence intervals. However, given the conservative nature of our calculations we can expect that the true range will be close to our estimated range.

Next, while it is very likely that currently unquantified detector effects will surface during running we do not anticipate any such effects posing a threat to the goals of this experiment. Mu3e has currently undergone several instrumentation and calibration runs and plans to prepare data taking in the near future. Thus far there has been no evidence of detector effects that present a significant deviation from what is expected for the lepton flavor violation experiment \citep{perrevoort2018searching}. Given the similarity between the requirements for the flavor violation experiment and our dark photon hunt, and the current progress made by the Mu3e experiment we do not anticipate any significant threat to our abilities to complete the goals of this experiment

The strongest challenge that this experiment will face is in performing the invariant mass reconstruction/vertex reconstruction. This will be a computationally intensive task given the large event numbers and will require the use of dedicated hardware. Correspondence with computer scientists will be required to discuss the complexity class of the task, as well as potential methods to speed up throughput (batching, local approximations etc.). Given the combinatorial nature of pairing the $e^{+}e^{-}$ pairs from detector events, this may be a prudent application for quantum combinatorial optimization. Further research and correspondence will be needed to determine the exact hardware and software required to achieve the goals of this proposal.

\section{Conclusion}

\begin{figure}
  \centering
  \includegraphics[scale=0.4]{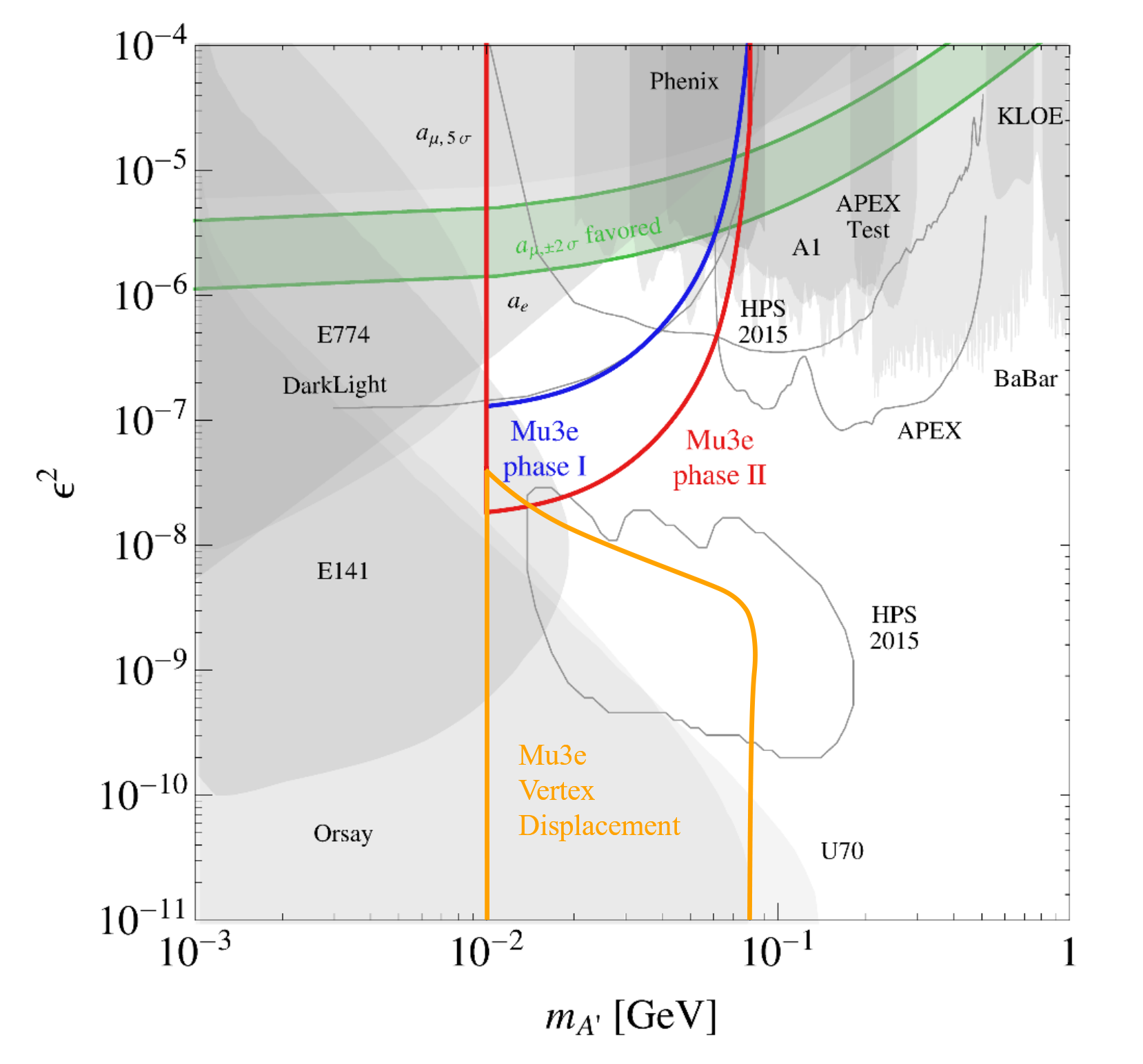}
  \centering
  \caption{Exclusion plots for the search. With the addition of the vertex displacement search the experiment will test a far larger parameter range than just the limits from the Atomki anomaly \citep{echenard2015projections}. Note that both the Orsay and U70 exclusions are $<90\%$ confidence intervals, generated from dark Higgs searches in the 1980's.}
  \label{fig:exclusion}
\end{figure}

In this proposal, we have motivated and outlined a new search for dark photons to be carried out at Mu3e. This search would specifically look for a dark photon within the parameter range of $10\text{ MeV}\leq m_{A'}\leq 80\text{ MeV}$ and $\epsilon\geq6\times10^{-7}$ using a resonance search combined with vertex reconstruction. As shown in Figure \ref{fig:exclusion}, the two approaches cover two intersecting regions, with the resonance search covering the entire range suggested by the Atomki anomaly. The vertex detection search will cover the lower segment of the $\epsilon$ parameter range. While this search appears to lie primarily in excluded regions of the Orsay and U70 experiments, it is critical to note that both of these experiments performed in the 1980s provided $<90\%$ confidence intervals for exclusion \citep{ilten2018serendipity}. The vertex detection search will then allow us to test and validate these limits to a higher confidence interval, while also exploring a new unexplored region of parameter space beyond the sensitivity of the HPS experiment \citep{celentano2014heavy}.

The results of this experiment will determine whether or not a dark photon exists to a high confidence interval within the cited parameter range, and whether an observed $A'$ corresponds with a dark photon corresponding to a novel $U(1)$ gauge symmetry as conjectured by Atomki and Feng et al. \citep{krasznahorkay2016observation, krasznahorkay2019new, feng2017particle}. The full theoretical implications of such a result have not yet been explored by this proposal’s authors. It will be a task for future papers and potentially other groups to determine what consequences such a result will have for physics beyond the Standard Model. Furthermore, the risks associated with running a research team and managing this project require robust consideration, and will require further planning before such a project breaks ground.

\subsection{Future Avenues}
The Mu3e experiment at PSI is adjacent to a low-temperature spallation neutron source (SINQ) connected to the same beam source. This could serve as a potential route for doing fixed-target $e-n$ or $\mu-n$ scattering. If a dark photon with the properties suggested by Atomki is observed, Mu3e would serve as an optimal setup for further experimentation and analysis of its properties, including whether it satisfies a $U(1)_B$ or $U(1)_{B-L}$ symmetry as suggested by Feng et al. (2017). Beyond the standard t-channel scattering process, whose matrix elements could be measured in the decay width of output products, the neutron has a magnetic dipole moment and may exhibit a Z-boson coupling. Matching the kinematic distributions of output products would enable calculation of the relative decay process matrix elements and elucidate the coupling of such a dark photon to the Standard Model \citep{jordan_smolinsky}.

\subsection{Acknowledgements}

The authors would like to deeply thank Jannicke Pearkes, Sam Carman, and David Goldhaber-Gordon for their many thoughtful suggestions and helpful feedback during the writing process. We are also deeply grateful for discussion and comments from Jordan Smolinsky regarding $A’$ gauge symmetries and potential background signals from neutron-electron scattering processes. Finally, we would like to thank Cameron Bravo for information on ongoing detector experiments and current exclusion regions for different dark photon candidates.
\newpage
\bibliography{citations}

\providecommand{\noopsort}[1]{}\providecommand{\singleletter}[1]{#1}%
\begin{thebibliography}{23}%
\makeatletter
\providecommand \@ifxundefined [1]{%
 \@ifx{#1\undefined}
}%
\providecommand \@ifnum [1]{%
 \ifnum #1\expandafter \@firstoftwo
 \else \expandafter \@secondoftwo
 \fi
}%
\providecommand \@ifx [1]{%
 \ifx #1\expandafter \@firstoftwo
 \else \expandafter \@secondoftwo
 \fi
}%
\providecommand \natexlab [1]{#1}%
\providecommand \enquote  [1]{``#1''}%
\providecommand \bibnamefont  [1]{#1}%
\providecommand \bibfnamefont [1]{#1}%
\providecommand \citenamefont [1]{#1}%
\providecommand \href@noop [0]{\@secondoftwo}%
\providecommand \href [0]{\begingroup \@sanitize@url \@href}%
\providecommand \@href[1]{\@@startlink{#1}\@@href}%
\providecommand \@@href[1]{\endgroup#1\@@endlink}%
\providecommand \@sanitize@url [0]{\catcode `\\12\catcode `\$12\catcode
  `\&12\catcode `\#12\catcode `\^12\catcode `\_12\catcode `\%12\relax}%
\providecommand \@@startlink[1]{}%
\providecommand \@@endlink[0]{}%
\providecommand \url  [0]{\begingroup\@sanitize@url \@url }%
\providecommand \@url [1]{\endgroup\@href {#1}{\urlprefix }}%
\providecommand \urlprefix  [0]{URL }%
\providecommand \Eprint [0]{\href }%
\providecommand \doibase [0]{http://dx.doi.org/}%
\providecommand \selectlanguage [0]{\@gobble}%
\providecommand \bibinfo  [0]{\@secondoftwo}%
\providecommand \bibfield  [0]{\@secondoftwo}%
\providecommand \translation [1]{[#1]}%
\providecommand \BibitemOpen [0]{}%
\providecommand \bibitemStop [0]{}%
\providecommand \bibitemNoStop [0]{.\EOS\space}%
\providecommand \EOS [0]{\spacefactor3000\relax}%
\providecommand \BibitemShut  [1]{\csname bibitem#1\endcsname}%
\let\auto@bib@innerbib\@empty
\bibitem [{\citenamefont {Krasznahorkay}\ \emph {et~al.}(2016)\citenamefont
  {Krasznahorkay}, \citenamefont {Csatl{\'o}s}, \citenamefont {Csige},
  \citenamefont {G{\'a}csi}, \citenamefont {Guly{\'a}s}, \citenamefont
  {Hunyadi}, \citenamefont {Kuti}, \citenamefont {Nyak{\'o}}, \citenamefont
  {Stuhl}, \citenamefont {Tim{\'a}r} \emph
  {et~al.}}]{krasznahorkay2016observation}%
  \BibitemOpen
  \bibfield  {author} {\bibinfo {author} {\bibfnamefont {A.}~\bibnamefont
  {Krasznahorkay}}, \bibinfo {author} {\bibfnamefont {M.}~\bibnamefont
  {Csatl{\'o}s}}, \bibinfo {author} {\bibfnamefont {L.}~\bibnamefont {Csige}},
  \bibinfo {author} {\bibfnamefont {Z.}~\bibnamefont {G{\'a}csi}}, \bibinfo
  {author} {\bibfnamefont {J.}~\bibnamefont {Guly{\'a}s}}, \bibinfo {author}
  {\bibfnamefont {M.}~\bibnamefont {Hunyadi}}, \bibinfo {author} {\bibfnamefont
  {I.}~\bibnamefont {Kuti}}, \bibinfo {author} {\bibfnamefont {B.}~\bibnamefont
  {Nyak{\'o}}}, \bibinfo {author} {\bibfnamefont {L.}~\bibnamefont {Stuhl}},
  \bibinfo {author} {\bibfnamefont {J.}~\bibnamefont {Tim{\'a}r}},  \emph
  {et~al.},\ }\href@noop {} {\bibfield  {journal} {\bibinfo  {journal}
  {Physical review letters}\ }\textbf {\bibinfo {volume} {116}},\ \bibinfo
  {pages} {042501} (\bibinfo {year} {2016})}\BibitemShut {NoStop}%
\bibitem [{\citenamefont {Krasznahorkay}\ \emph {et~al.}(2019)\citenamefont
  {Krasznahorkay}, \citenamefont {Csatl{\'o}s}, \citenamefont {Csige},
  \citenamefont {Gulyas}, \citenamefont {Koszta}, \citenamefont {Szihalmi},
  \citenamefont {Tim{\'a}r}, \citenamefont {Firak}, \citenamefont {Nagy},
  \citenamefont {Sas} \emph {et~al.}}]{krasznahorkay2019new}%
  \BibitemOpen
  \bibfield  {author} {\bibinfo {author} {\bibfnamefont {A.}~\bibnamefont
  {Krasznahorkay}}, \bibinfo {author} {\bibfnamefont {M.}~\bibnamefont
  {Csatl{\'o}s}}, \bibinfo {author} {\bibfnamefont {L.}~\bibnamefont {Csige}},
  \bibinfo {author} {\bibfnamefont {J.}~\bibnamefont {Gulyas}}, \bibinfo
  {author} {\bibfnamefont {M.}~\bibnamefont {Koszta}}, \bibinfo {author}
  {\bibfnamefont {B.}~\bibnamefont {Szihalmi}}, \bibinfo {author}
  {\bibfnamefont {J.}~\bibnamefont {Tim{\'a}r}}, \bibinfo {author}
  {\bibfnamefont {D.}~\bibnamefont {Firak}}, \bibinfo {author} {\bibfnamefont
  {A.}~\bibnamefont {Nagy}}, \bibinfo {author} {\bibfnamefont {N.}~\bibnamefont
  {Sas}},  \emph {et~al.},\ }\href@noop {} {\bibfield  {journal} {\bibinfo
  {journal} {arXiv preprint arXiv:1910.10459}\ } (\bibinfo {year}
  {2019})}\BibitemShut {NoStop}%
\bibitem [{\citenamefont {Feng}\ \emph {et~al.}(2017)\citenamefont {Feng},
  \citenamefont {Fornal}, \citenamefont {Galon}, \citenamefont {Gardner},
  \citenamefont {Smolinsky}, \citenamefont {Tait},\ and\ \citenamefont
  {Tanedo}}]{feng2017particle}%
  \BibitemOpen
  \bibfield  {author} {\bibinfo {author} {\bibfnamefont {J.~L.}\ \bibnamefont
  {Feng}}, \bibinfo {author} {\bibfnamefont {B.}~\bibnamefont {Fornal}},
  \bibinfo {author} {\bibfnamefont {I.}~\bibnamefont {Galon}}, \bibinfo
  {author} {\bibfnamefont {S.}~\bibnamefont {Gardner}}, \bibinfo {author}
  {\bibfnamefont {J.}~\bibnamefont {Smolinsky}}, \bibinfo {author}
  {\bibfnamefont {T.~M.}\ \bibnamefont {Tait}}, \ and\ \bibinfo {author}
  {\bibfnamefont {P.}~\bibnamefont {Tanedo}},\ }\href@noop {} {\bibfield
  {journal} {\bibinfo  {journal} {Physical Review D}\ }\textbf {\bibinfo
  {volume} {95}},\ \bibinfo {pages} {035017} (\bibinfo {year}
  {2017})}\BibitemShut {NoStop}%
\bibitem [{\citenamefont {Alves}\ and\ \citenamefont
  {Weiner}(2018)}]{alves2018viable}%
  \BibitemOpen
  \bibfield  {author} {\bibinfo {author} {\bibfnamefont {D.~S.}\ \bibnamefont
  {Alves}}\ and\ \bibinfo {author} {\bibfnamefont {N.}~\bibnamefont {Weiner}},\
  }\href@noop {} {\bibfield  {journal} {\bibinfo  {journal} {Journal of High
  Energy Physics}\ }\textbf {\bibinfo {volume} {2018}},\ \bibinfo {pages} {92}
  (\bibinfo {year} {2018})}\BibitemShut {NoStop}%
\bibitem [{\citenamefont {Bennett}\ \emph {et~al.}(2006)\citenamefont
  {Bennett}, \citenamefont {Bousquet}, \citenamefont {Brown}, \citenamefont
  {Bunce}, \citenamefont {Carey}, \citenamefont {Cushman}, \citenamefont
  {Danby}, \citenamefont {Debevec}, \citenamefont {Deile}, \citenamefont {Deng}
  \emph {et~al.}}]{bennett2006final}%
  \BibitemOpen
  \bibfield  {author} {\bibinfo {author} {\bibfnamefont {G.~W.}\ \bibnamefont
  {Bennett}}, \bibinfo {author} {\bibfnamefont {B.}~\bibnamefont {Bousquet}},
  \bibinfo {author} {\bibfnamefont {H.}~\bibnamefont {Brown}}, \bibinfo
  {author} {\bibfnamefont {G.}~\bibnamefont {Bunce}}, \bibinfo {author}
  {\bibfnamefont {R.}~\bibnamefont {Carey}}, \bibinfo {author} {\bibfnamefont
  {P.}~\bibnamefont {Cushman}}, \bibinfo {author} {\bibfnamefont
  {G.}~\bibnamefont {Danby}}, \bibinfo {author} {\bibfnamefont
  {P.}~\bibnamefont {Debevec}}, \bibinfo {author} {\bibfnamefont
  {M.}~\bibnamefont {Deile}}, \bibinfo {author} {\bibfnamefont
  {H.}~\bibnamefont {Deng}},  \emph {et~al.},\ }\href@noop {} {\bibfield
  {journal} {\bibinfo  {journal} {Physical Review D}\ }\textbf {\bibinfo
  {volume} {73}},\ \bibinfo {pages} {072003} (\bibinfo {year}
  {2006})}\BibitemShut {NoStop}%
\bibitem [{\citenamefont {Carone}(2013)}]{carone2013flavor}%
  \BibitemOpen
  \bibfield  {author} {\bibinfo {author} {\bibfnamefont {C.~D.}\ \bibnamefont
  {Carone}},\ }\href@noop {} {\bibfield  {journal} {\bibinfo  {journal}
  {Physics Letters B}\ }\textbf {\bibinfo {volume} {721}},\ \bibinfo {pages}
  {118} (\bibinfo {year} {2013})}\BibitemShut {NoStop}%
\bibitem [{\citenamefont {Fradette}\ \emph {et~al.}(2014)\citenamefont
  {Fradette}, \citenamefont {Pospelov}, \citenamefont {Pradler},\ and\
  \citenamefont {Ritz}}]{fradette2014cosmological}%
  \BibitemOpen
  \bibfield  {author} {\bibinfo {author} {\bibfnamefont {A.}~\bibnamefont
  {Fradette}}, \bibinfo {author} {\bibfnamefont {M.}~\bibnamefont {Pospelov}},
  \bibinfo {author} {\bibfnamefont {J.}~\bibnamefont {Pradler}}, \ and\
  \bibinfo {author} {\bibfnamefont {A.}~\bibnamefont {Ritz}},\ }\href@noop {}
  {\bibfield  {journal} {\bibinfo  {journal} {Physical Review D}\ }\textbf
  {\bibinfo {volume} {90}},\ \bibinfo {pages} {035022} (\bibinfo {year}
  {2014})}\BibitemShut {NoStop}%
\bibitem [{\citenamefont {Beacham}(2014)}]{beacham2014apex}%
  \BibitemOpen
  \bibfield  {author} {\bibinfo {author} {\bibfnamefont {J.}~\bibnamefont
  {Beacham}},\ }in\ \href@noop {} {\emph {\bibinfo {booktitle} {EPJ Web of
  Conferences}}},\ Vol.~\bibinfo {volume} {73}\ (\bibinfo {organization} {EDP
  Sciences},\ \bibinfo {year} {2014})\ p.\ \bibinfo {pages} {07011}\BibitemShut
  {NoStop}%
\bibitem [{\citenamefont {Guly{\'a}s}\ \emph {et~al.}(2016)\citenamefont
  {Guly{\'a}s}, \citenamefont {Ketel}, \citenamefont {Krasznahorkay},
  \citenamefont {Csatl{\'o}s}, \citenamefont {Csige}, \citenamefont
  {G{\'a}csi}, \citenamefont {Hunyadi}, \citenamefont {Krasznahorkay},
  \citenamefont {Vit{\'e}z},\ and\ \citenamefont {Tornyi}}]{gulyas2016pair}%
  \BibitemOpen
  \bibfield  {author} {\bibinfo {author} {\bibfnamefont {J.}~\bibnamefont
  {Guly{\'a}s}}, \bibinfo {author} {\bibfnamefont {T.}~\bibnamefont {Ketel}},
  \bibinfo {author} {\bibfnamefont {A.}~\bibnamefont {Krasznahorkay}}, \bibinfo
  {author} {\bibfnamefont {M.}~\bibnamefont {Csatl{\'o}s}}, \bibinfo {author}
  {\bibfnamefont {L.}~\bibnamefont {Csige}}, \bibinfo {author} {\bibfnamefont
  {Z.}~\bibnamefont {G{\'a}csi}}, \bibinfo {author} {\bibfnamefont
  {M.}~\bibnamefont {Hunyadi}}, \bibinfo {author} {\bibfnamefont
  {A.}~\bibnamefont {Krasznahorkay}}, \bibinfo {author} {\bibfnamefont
  {A.}~\bibnamefont {Vit{\'e}z}}, \ and\ \bibinfo {author} {\bibfnamefont
  {T.}~\bibnamefont {Tornyi}},\ }\href@noop {} {\bibfield  {journal} {\bibinfo
  {journal} {Nuclear Instruments and Methods in Physics Research Section A:
  Accelerators, Spectrometers, Detectors and Associated Equipment}\ }\textbf
  {\bibinfo {volume} {808}},\ \bibinfo {pages} {21} (\bibinfo {year}
  {2016})}\BibitemShut {NoStop}%
\bibitem [{\citenamefont {Banerjee}\ \emph {et~al.}(2018)\citenamefont
  {Banerjee}, \citenamefont {Burtsev}, \citenamefont {Chumakov}, \citenamefont
  {Cooke}, \citenamefont {Crivelli}, \citenamefont {Depero}, \citenamefont
  {Dermenev}, \citenamefont {Donskov}, \citenamefont {Dusaev}, \citenamefont
  {Enik} \emph {et~al.}}]{banerjee2018search}%
  \BibitemOpen
  \bibfield  {author} {\bibinfo {author} {\bibfnamefont {D.}~\bibnamefont
  {Banerjee}}, \bibinfo {author} {\bibfnamefont {V.~E.}\ \bibnamefont
  {Burtsev}}, \bibinfo {author} {\bibfnamefont {A.~G.}\ \bibnamefont
  {Chumakov}}, \bibinfo {author} {\bibfnamefont {D.}~\bibnamefont {Cooke}},
  \bibinfo {author} {\bibfnamefont {P.}~\bibnamefont {Crivelli}}, \bibinfo
  {author} {\bibfnamefont {E.}~\bibnamefont {Depero}}, \bibinfo {author}
  {\bibfnamefont {A.}~\bibnamefont {Dermenev}}, \bibinfo {author}
  {\bibfnamefont {S.}~\bibnamefont {Donskov}}, \bibinfo {author} {\bibfnamefont
  {R.}~\bibnamefont {Dusaev}}, \bibinfo {author} {\bibfnamefont
  {T.}~\bibnamefont {Enik}},  \emph {et~al.},\ }\href@noop {} {\bibfield
  {journal} {\bibinfo  {journal} {Physical review letters}\ }\textbf {\bibinfo
  {volume} {120}},\ \bibinfo {pages} {231802} (\bibinfo {year}
  {2018})}\BibitemShut {NoStop}%
\bibitem [{\citenamefont {Wojtsekhowski}\ \emph {et~al.}(2018)\citenamefont
  {Wojtsekhowski}, \citenamefont {Baranov}, \citenamefont {Blinov},
  \citenamefont {Levichev}, \citenamefont {Mishnev}, \citenamefont {Nikolenko},
  \citenamefont {Rachek}, \citenamefont {Shestakov}, \citenamefont {Tikhonov},
  \citenamefont {Toporkov} \emph {et~al.}}]{wojtsekhowski2018searching}%
  \BibitemOpen
  \bibfield  {author} {\bibinfo {author} {\bibfnamefont {B.}~\bibnamefont
  {Wojtsekhowski}}, \bibinfo {author} {\bibfnamefont {G.}~\bibnamefont
  {Baranov}}, \bibinfo {author} {\bibfnamefont {M.}~\bibnamefont {Blinov}},
  \bibinfo {author} {\bibfnamefont {E.}~\bibnamefont {Levichev}}, \bibinfo
  {author} {\bibfnamefont {S.}~\bibnamefont {Mishnev}}, \bibinfo {author}
  {\bibfnamefont {D.}~\bibnamefont {Nikolenko}}, \bibinfo {author}
  {\bibfnamefont {I.}~\bibnamefont {Rachek}}, \bibinfo {author} {\bibfnamefont
  {Y.~V.}\ \bibnamefont {Shestakov}}, \bibinfo {author} {\bibfnamefont {Y.~A.}\
  \bibnamefont {Tikhonov}}, \bibinfo {author} {\bibfnamefont {D.}~\bibnamefont
  {Toporkov}},  \emph {et~al.},\ }\href@noop {} {\bibfield  {journal} {\bibinfo
   {journal} {Journal of Instrumentation}\ }\textbf {\bibinfo {volume} {13}},\
  \bibinfo {pages} {P02021} (\bibinfo {year} {2018})}\BibitemShut {NoStop}%
\bibitem [{\citenamefont {Katzin}(2012)}]{katzin2012darklight}%
  \BibitemOpen
  \bibfield  {author} {\bibinfo {author} {\bibfnamefont {D.~R.}\ \bibnamefont
  {Katzin}},\ }\emph {\bibinfo {title} {The DarkLight experiment: searching for
  the dark photon}},\ \href@noop {} {Ph.D. thesis},\ \bibinfo  {school}
  {Massachusetts Institute of Technology} (\bibinfo {year} {2012})\BibitemShut
  {NoStop}%
\bibitem [{\citenamefont {Echenard}\ \emph {et~al.}(2015)\citenamefont
  {Echenard}, \citenamefont {Essig},\ and\ \citenamefont
  {Zhong}}]{echenard2015projections}%
  \BibitemOpen
  \bibfield  {author} {\bibinfo {author} {\bibfnamefont {B.}~\bibnamefont
  {Echenard}}, \bibinfo {author} {\bibfnamefont {R.}~\bibnamefont {Essig}}, \
  and\ \bibinfo {author} {\bibfnamefont {Y.-M.}\ \bibnamefont {Zhong}},\
  }\href@noop {} {\bibfield  {journal} {\bibinfo  {journal} {Journal of High
  Energy Physics}\ }\textbf {\bibinfo {volume} {2015}},\ \bibinfo {pages} {113}
  (\bibinfo {year} {2015})}\BibitemShut {NoStop}%
\bibitem [{\citenamefont {Fabbrichesi}\ \emph {et~al.}(2020)\citenamefont
  {Fabbrichesi}, \citenamefont {Gabrielli},\ and\ \citenamefont
  {Lanfranchi}}]{fabbrichesi2020dark}%
  \BibitemOpen
  \bibfield  {author} {\bibinfo {author} {\bibfnamefont {M.}~\bibnamefont
  {Fabbrichesi}}, \bibinfo {author} {\bibfnamefont {E.}~\bibnamefont
  {Gabrielli}}, \ and\ \bibinfo {author} {\bibfnamefont {G.}~\bibnamefont
  {Lanfranchi}},\ }\href@noop {} {\bibfield  {journal} {\bibinfo  {journal}
  {arXiv preprint arXiv:2005.01515}\ } (\bibinfo {year} {2020})}\BibitemShut
  {NoStop}%
\bibitem [{\citenamefont {Watkins}(1986)}]{watkins1986discovery}%
  \BibitemOpen
  \bibfield  {author} {\bibinfo {author} {\bibfnamefont {P.~M.}\ \bibnamefont
  {Watkins}},\ }\href@noop {} {\bibfield  {journal} {\bibinfo  {journal}
  {Contemporary Physics}\ }\textbf {\bibinfo {volume} {27}},\ \bibinfo {pages}
  {291} (\bibinfo {year} {1986})}\BibitemShut {NoStop}%
\bibitem [{\citenamefont {Berger}\ \emph {et~al.}(2014)\citenamefont {Berger},
  \citenamefont {Collaboration} \emph {et~al.}}]{berger2014mu3e}%
  \BibitemOpen
  \bibfield  {author} {\bibinfo {author} {\bibfnamefont {N.}~\bibnamefont
  {Berger}}, \bibinfo {author} {\bibfnamefont {M.}~\bibnamefont
  {Collaboration}},  \emph {et~al.},\ }\href@noop {} {\bibfield  {journal}
  {\bibinfo  {journal} {Nuclear Physics B-Proceedings Supplements}\ }\textbf
  {\bibinfo {volume} {248}},\ \bibinfo {pages} {35} (\bibinfo {year}
  {2014})}\BibitemShut {NoStop}%
\bibitem [{\citenamefont {Blondel}\ \emph {et~al.}(2013)\citenamefont
  {Blondel}, \citenamefont {Bravar}, \citenamefont {Pohl}, \citenamefont
  {Bachmann}, \citenamefont {Berger}, \citenamefont {Kiehn}, \citenamefont
  {Sch{\"o}ning}, \citenamefont {Wiedner}, \citenamefont {Windelband},
  \citenamefont {Eckert} \emph {et~al.}}]{blondel2013research}%
  \BibitemOpen
  \bibfield  {author} {\bibinfo {author} {\bibfnamefont {A.}~\bibnamefont
  {Blondel}}, \bibinfo {author} {\bibfnamefont {A.}~\bibnamefont {Bravar}},
  \bibinfo {author} {\bibfnamefont {M.}~\bibnamefont {Pohl}}, \bibinfo {author}
  {\bibfnamefont {S.}~\bibnamefont {Bachmann}}, \bibinfo {author}
  {\bibfnamefont {N.}~\bibnamefont {Berger}}, \bibinfo {author} {\bibfnamefont
  {M.}~\bibnamefont {Kiehn}}, \bibinfo {author} {\bibfnamefont
  {A.}~\bibnamefont {Sch{\"o}ning}}, \bibinfo {author} {\bibfnamefont
  {D.}~\bibnamefont {Wiedner}}, \bibinfo {author} {\bibfnamefont
  {B.}~\bibnamefont {Windelband}}, \bibinfo {author} {\bibfnamefont
  {P.}~\bibnamefont {Eckert}},  \emph {et~al.},\ }\href@noop {} {\bibfield
  {journal} {\bibinfo  {journal} {arXiv preprint arXiv:1301.6113}\ } (\bibinfo
  {year} {2013})}\BibitemShut {NoStop}%
\bibitem [{\citenamefont {Augustin}\ \emph {et~al.}(2019)\citenamefont
  {Augustin}, \citenamefont {Berger}, \citenamefont {Dittmeier}, \citenamefont
  {Ehrler}, \citenamefont {Grzesik}, \citenamefont {Hammerich}, \citenamefont
  {Herkert}, \citenamefont {Huth}, \citenamefont {Kr{\"o}ger}, \citenamefont
  {Aeschbacher} \emph {et~al.}}]{augustin2019mupix8}%
  \BibitemOpen
  \bibfield  {author} {\bibinfo {author} {\bibfnamefont {H.}~\bibnamefont
  {Augustin}}, \bibinfo {author} {\bibfnamefont {N.}~\bibnamefont {Berger}},
  \bibinfo {author} {\bibfnamefont {S.}~\bibnamefont {Dittmeier}}, \bibinfo
  {author} {\bibfnamefont {F.}~\bibnamefont {Ehrler}}, \bibinfo {author}
  {\bibfnamefont {C.}~\bibnamefont {Grzesik}}, \bibinfo {author} {\bibfnamefont
  {J.}~\bibnamefont {Hammerich}}, \bibinfo {author} {\bibfnamefont
  {A.}~\bibnamefont {Herkert}}, \bibinfo {author} {\bibfnamefont
  {L.}~\bibnamefont {Huth}}, \bibinfo {author} {\bibfnamefont {J.}~\bibnamefont
  {Kr{\"o}ger}}, \bibinfo {author} {\bibfnamefont {F.~M.}\ \bibnamefont
  {Aeschbacher}},  \emph {et~al.},\ }\href@noop {} {\bibfield  {journal}
  {\bibinfo  {journal} {Nuclear Instruments and Methods in Physics Research
  Section A: Accelerators, Spectrometers, Detectors and Associated Equipment}\
  }\textbf {\bibinfo {volume} {936}},\ \bibinfo {pages} {681} (\bibinfo {year}
  {2019})}\BibitemShut {NoStop}%
\bibitem [{\citenamefont {Kuno}\ and\ \citenamefont
  {Okada}(2001)}]{kuno2001muon}%
  \BibitemOpen
  \bibfield  {author} {\bibinfo {author} {\bibfnamefont {Y.}~\bibnamefont
  {Kuno}}\ and\ \bibinfo {author} {\bibfnamefont {Y.}~\bibnamefont {Okada}},\
  }\href@noop {} {\bibfield  {journal} {\bibinfo  {journal} {Reviews of Modern
  Physics}\ }\textbf {\bibinfo {volume} {73}},\ \bibinfo {pages} {151}
  (\bibinfo {year} {2001})}\BibitemShut {NoStop}%
\bibitem [{\citenamefont {Perrevoort}(2018)}]{perrevoort2018searching}%
  \BibitemOpen
  \bibfield  {author} {\bibinfo {author} {\bibfnamefont {A.-K.}\ \bibnamefont
  {Perrevoort}},\ }\href@noop {} {\bibfield  {journal} {\bibinfo  {journal}
  {arXiv preprint arXiv:1802.09851}\ } (\bibinfo {year} {2018})}\BibitemShut
  {NoStop}%
\bibitem [{\citenamefont {Ilten}\ \emph {et~al.}(2018)\citenamefont {Ilten},
  \citenamefont {Soreq}, \citenamefont {Williams},\ and\ \citenamefont
  {Xue}}]{ilten2018serendipity}%
  \BibitemOpen
  \bibfield  {author} {\bibinfo {author} {\bibfnamefont {P.}~\bibnamefont
  {Ilten}}, \bibinfo {author} {\bibfnamefont {Y.}~\bibnamefont {Soreq}},
  \bibinfo {author} {\bibfnamefont {M.}~\bibnamefont {Williams}}, \ and\
  \bibinfo {author} {\bibfnamefont {W.}~\bibnamefont {Xue}},\ }\href@noop {}
  {\bibfield  {journal} {\bibinfo  {journal} {Journal of High Energy Physics}\
  }\textbf {\bibinfo {volume} {2018}},\ \bibinfo {pages} {4} (\bibinfo {year}
  {2018})}\BibitemShut {NoStop}%
\bibitem [{\citenamefont {Celentano}\ \emph {et~al.}(2014)\citenamefont
  {Celentano}, \citenamefont {Collaboration} \emph
  {et~al.}}]{celentano2014heavy}%
  \BibitemOpen
  \bibfield  {author} {\bibinfo {author} {\bibfnamefont {A.}~\bibnamefont
  {Celentano}}, \bibinfo {author} {\bibfnamefont {H.}~\bibnamefont
  {Collaboration}},  \emph {et~al.},\ }in\ \href@noop {} {\emph {\bibinfo
  {booktitle} {Journal of Physics: Conference Series}}},\ Vol.\ \bibinfo
  {volume} {556}\ (\bibinfo {organization} {IOP Publishing},\ \bibinfo {year}
  {2014})\ p.\ \bibinfo {pages} {012064}\BibitemShut {NoStop}%
\bibitem [{\citenamefont {Smolinsky}(2020)}]{jordan_smolinsky}%
  \BibitemOpen
  \bibfield  {author} {\bibinfo {author} {\bibfnamefont {J.}~\bibnamefont
  {Smolinsky}},\ }\href@noop {} {}\bibinfo {howpublished} {{Private
  correspondence}} (\bibinfo {year} {2020})\BibitemShut {NoStop}%
\end{thebibliography}%
\end{document}